\documentclass[aps,prd,twocolumn,preprintnumbers,floatfix,nofootinbib]{revtex4}

\usepackage[usenames]{color}
\usepackage{subfigure,amsmath, amssymb ,amsfonts, latexsym}  
\usepackage[final]{graphicx}
\usepackage[colorlinks,hyperindex, 
            bookmarks=true,bookmarksopen=true]{hyperref}
    \hypersetup
    { 
        colorlinks=true,
        linkcolor=blue,
        urlcolor=blue,
        filecolor=black,
        citecolor=red,
        pdfstartview=FitV,
        pdftitle={},
        pdfauthor={Nick Evans, Keun-Young Kim, Maria Magou, Andy O'Bannon},
        pdfsubject={},
        pdfkeywords={},
        pdfpagemode=None,
        bookmarksopen=true
    }

\usepackage[notcite, color, final
                               ]{showkeys}
\definecolor{refkey}{gray}{.5}
\definecolor{labelkey}{gray}{.5}

       \def\e  {\epsilon}

\renewcommand{\a}{\alpha}      \renewcommand{\b}{\beta}
      
     \renewcommand{\L}{\Lambda}

 \newcommand{\call}{\mbox{${\cal L}$}}
 \newcommand{\caln}{\mbox{${\cal N}$}}

\newcommand{\be}{\begin{equation}}
\newcommand{\ee}{\end{equation}}
\newcommand{\beqa}{\begin{subequations}\begin{eqnarray}}
\newcommand{\eeqa}{\end{eqnarray}\end{subequations}}

\newcommand{\ra}{\rightarrow}

\newcommand{\ud}{\,\mathrm{d}}
\newcommand{\dd}{\mathrm{d}}
\newcommand{\del}{\partial}

\begin{document}

\newcommand\sect[1]{\emph{#1}---}

\preprint{
\begin{minipage}[t]{3in}
\begin{flushright} SHEP-11-16
\\[30pt]
\hphantom{.}
\end{flushright}
\end{minipage}
}

\title{Holographic Wilsonian Renormalization and Chiral Phase Transitions}

\author{Nick Evans}
\email{evans@soton.ac.uk}
\author{Keun-Young Kim}
\email{k.kim@soton.ac.uk}
\author{Maria Magou}
\email{mm21g08@soton.ac.uk}
\affiliation{ School of Physics and Astronomy, University of
Southampton, Southampton, SO17 1BJ, UK 
}

\begin{abstract}
\noindent We explore the role of a holographic Wilsonian cut-off in simple probe brane
models with chiral symmetry 
breaking/restoration phase transitions. The Wilsonian cut-off allows us to define supergravity
solutions for off-shell configurations and hence to define a potential for the chiral condensate.
We pay particular attention to the need for configurations whose action we are comparing to have
the same IR and UV boundary conditions.
We exhibit new first and second order phase transitions with changing cut-off. We derive 
the effective potential for the condensate 
including mean field and BKT type
continuous transitions.

\end{abstract}

\maketitle


\section{Introduction}

Recently a number of authors have proposed using renormalization ideas in the spirit
of Wilson in holographic descriptions of strongly
coupled gauge theories~\cite{Bredberg:2010ky,Nickel:2010pr,Heemskerk:2010hk,Faulkner:2010jy,Douglas:2010rc,Sin:2011yh,Lee:2009ij,Lee:2010ub}.
(For earlier discussions, 
see~\cite{Akhmedov:1998vf,Akhmedov:2002gq,deBoer:1999xf,deBoer:2000cz,Lewandowski:2002rf,Lewandowski:2004yr})
The radial direction in AdS like spaces is dual to energy
scale in the field theory~\cite{Maldacena:1997re,Susskind:1998dq,Peet:1998wn,Evans:2006eq} 
and one can imagine introducing a cut-off at some finite
radius, splitting the supergravity solution in two. By integrating out the high energy regime an effective Wilsonian description should emerge. The precise matching of
the radial direction to a gauge invariant measure of energy scale remains an open problem
so a precise match to Wilsonian renormalization in the field theory will also remain imprecise
but the spirit is clear.

In this paper we wish to bring these ideas to bare on some explicit examples of theories
with phase transitions. We wish to study how those transitions emerge in the Wilsonian
language and will find examples of new transitions with changing Wilsonian cut-off scale. We are also
interested in deriving low energy effective actions near the transition points using this
language.

In particular we will use the simple but highly instructive 
D3/D7 and D3/D5 systems~\cite{Karch:2002sh}. The D3/D7 theory is the
${\cal N}=4$ Super Yang-Mills theory in 3+1 dimension with ${\cal N}=2$ quark hypermultiplets. In
the D3/D5 case the hypermultiplets are restricted to a 2+1 dimension sub-surface of the
gauge theory. In the
quenched limit, when the number of quark flavours is small but the number of colours large,
we can use the gauge/gravity dual consisting of probe D7(D5) branes in AdS$_5\times$S$^5$. This
system has been widely explored (For example in the D3/D7 case see \cite{Erdmenger:2007cm,Babington:2003vm,Nakamura:2006xk,Kobayashi:2006sb,Filev:2007gb,Karch:2007pd,Evans:2010iy,Jensen:2010vd,Evans:2011mu,Evans:2010np,Evans:2010xs,Evans:2010tf}  and references therein and \cite{Karch:2000gx,DeWolfe:2001pq,Erdmenger:2002ex,Myers:2008me,Jensen:2010ga,Evans:2010hi} for the D3/D5 case) 
and we will make use of a number of known phenomena. In these systems the radial coordinate on the probe brane plays the role of the 
renormalization group (RG) scale in the field theory. 

We will first introduce our methodology in the supersymmetric ${\cal N}=2$ theory. That theory does
not induce a chiral quark condensate (which would break supersymmetry were it present) but we can nevertheless
attempt to find an effective potential for the quark condensate which should be minimized at zero. We
will study its dependence on changing Wilsonian scale. This introduces the first subtlety which is the need to define
holographic flows for non-vacuum, ``off-shell'', states in the field theory. In the UV, solutions of
the Euler Lagrange equation for the D7 embedding exist for all values of the condensate. 
In fact we show analytically in this case that 
the D7 embeddings with non-zero condensate become complex at some finite AdS radius. At any fixed radius the solutions
that are still real 
do not share the same boundary condition so formally one should not cut them off and compare their actions.
To remedy this we consider a cut-off with explicit width i.e. effectively two close cut-offs. 
We use the naive UV solutions down to the the higher cut-off but then match them to classical embeddings 
between the two cut-offs that share the same IR boundary conditions. 
After making this construction one
can then take the limit where the cut-offs come together. In this case that
limit leaves us just evaluating the UV flow's action down to the cut-off as one naively expects,
however it prepares the ground work for later more complicated cases. 
If the cut off is taken too low
then an embedding will become complex and computing the action is impossible. We interpret this as
high energy states being integrated from the low energy effective theory - these states have
energy above the cut off and are simply no longer present in the Wilsonian effective IR theory.

The precise meaning in the field 
theory of any cut-off we introduce of course is ambiguous but we presume there is some sensible
mapping. Indeed there are also many distinct ways in which a cut-off can be introduced
in the field theory from a sharp cut on UV modes to some smooth function suppressing the UV 
contributions. Using our prescription for the cut-off, we then evaluate
the action of the D7 brane, which is just
the free energy in the field theory. If we evaluate the UV component of that action above our cut-off
we are simply determining the effective classical potential for the quark condensate that encapsulates the
physics above that scale. This is the Wilsonian effective potential. The deep IR of this potential
only contains the vacuum state with the condensate equal to zero since all other states are
associated with complex embeddings - we can though freeze the energy of those states at the point
they are integrated out (become complex) to generate an IR effective potential.

In the presence of a magnetic field, $B$,in the D3/D7 system
a quark condensate is induced that breaks a $U(1)$ chiral
symmetry~\cite{Filev:2007gb}. In this system we again study the effective potential for the quark condensate with
changing Wilsonian scale using the ideas so far developed. 
In the pure $B$ case the resulting RG flow shows a novel second order transition to the
symmetry breaking configuration as the Wilsonian scale is changed. This is an example
of a strongly coupled Coleman Weinberg \cite{Coleman:1973jx} style symmetry breaking. We also holographically compute the
effective potential
close to the transition and show it is mean field in nature. We  plot 
the RG flow of the couplings of the Wilsonian effective potential.  In the deep IR the effective potential
for the condensate again develops gaps as embeddings become complex and are integrated from the low
energy theory.
The picture that emerges is satisfyingly Wilsonian. The bare UV theory has no symmetry breaking;
at intermediate RG scales integrated out, UV, quantum effects enter the bare potential and
display the symmetry breaking; in the deep IR all states but the true vacuum are integrated 
from the low energy theory.
The Wilsonian approach gives a sensible intuition.

The pure $B$ theory can also be related immediately to the case of a $B$ field and a perpendicular electric field, $E$
\cite{OBannon:2007in,Kim:2011qh,Evans:2011tk}.
These two theories essentially share the same action.
The $E$ field~\cite{Karch:2007pd,Erdmenger:2007bn,Albash:2007bq} tends to dissociate quark bound states 
and so disfavours chiral symmetry breaking~\cite{Evans:2011tk}. 
We show that our previous results can be mapped to display the $E$ dependence of the Wilsonian
description.

We next turn our techniques to analyze the D3/D7 system with a magnetic field and 
chemical potential~\cite{Evans:2010iy,Jensen:2010vd}.
The chemical potential tends to induce a non-zero quark density which also disrupts
the chiral condensate. Here the naive embedding flows for the D7 brane, describing
different condensate values, all progress to the deep IR where they mostly end in a singular
fashion. Previously those flows that end at the position of the D3 branes (the origin) have been picked
out to describe the physical vacuum~\cite{Kobayashi:2006sb}. The picture is that fundamental strings, representing
the quark density, link the D3 and D7 brane. They manifest as a spike in the D7 brane embedding
to the origin. The fundamental strings are needed to source the D7 brane world volume
gauge field that is dual to chemical potential. To compare the actions of these vacuum flows and the
off shell configurations, the off shell configurations must be forced to have the same IR boundary conditions.
We use our cut-off procedure to argue that in the deep IR the off-shell configuration should be 
completed with a spike of D7 brane to the origin. The natural extension of this procedure at 
non-zero cut-off values is that all configurations should be completed with a spike along the cut-off.

Having argued for this implementation of the cut-off we then analyze the Wilsonian effective potential
of this theory at fixed B but varying density. Again we find a sensible Wilsonian description
with the UV theory showing no symmetry breaking. Then, provided the density is sufficiently small,   
there is a transition with lowering cut-off scale to the chiral symmetry broken vacuum. This transition
is in parts of parameter space first order and elsewhere 
second order and mean field. We can explicitly derive the effective potential through the 
transition. As the cut-off is taken into the IR the second order behaviour dominates and 
we perform a fit to the mean field potential.
In this case we do not see the degeneracy of the potential in the deep
IR we described in the supersymmetric and pure B case - none of the embeddings become complex.
This might reflect that our cut-off prescription is overly 
naive. We simply report on what we find in this case.

Finally we study the D3/D5 system~\cite{Karch:2000gx,DeWolfe:2001pq,Erdmenger:2002ex} with a magnetic field and density, 
$d$, using our Wilsonian methodology. 
This system is of further interest because it is known to exhibit a holographic BKT transition~\cite{Jensen:2010ga,Evans:2010hi}
at which the condensate grows as $e^{-1/\sqrt{d_c-d}}$. Here we again display the
density versus cut-off phase diagram,
in which there are first order transition regimes, second order regimes
and finally for the cut-off in the deep IR a BKT transition. Here we 
successfully derive an effective potential for the BKT transitions when 
the Wilsonian scale goes to zero.

\section{Wilsonian Flow For the ${\cal N}=2$ Theory}

We will begin by exploring a Wilsonian analysis of the simplest model 
${\cal N}=2$ gauge theory which does not display chiral symmetry breaking.
The ${\cal N}=4$ gauge theory at zero temperature is described by the dual geometry (AdS$_5\times S^5$)
\begin{equation} 
\ud s^2 = \frac{r^2}{R^2} \ud x_4^2 +
\frac{R^2}{r^2} (\ud \rho^2 + \rho^2
\ud \Omega_3^2 + \ud L^2 + L^2 \ud \phi^2) \,,
\end{equation}
where $r^2= \rho^2 + L^2$ and $ R^4=4 \pi g_s N \alpha'^{2}$.

Quenched ($N_f \ll N$) ${\cal N}$=2 quark superfields can be
included through probe D7 branes
in the geometry.
The D3-D7 strings are the quarks. D7-D7 strings holographically
describe mesonic operators and their sources. The D7 probe can be
described by its DBI action
\begin{equation} \label{dbi1}
S_{\mathrm{DBI}} = - T_{D7} \int \ud^8\xi \sqrt{- {\rm det} (P[G]_{ab} +
2 \pi \alpha' F_{ab})} \ , 
\end{equation}
where $T_{D7} = (2\pi)^{-7}\a'^{-4} g_s^{-1}$ and $P[G]_{ab}$ is the pullback of the metric and $F_{ab}$ is
the gauge field living on the D7 world volume. 
The Wess-Zumino term is irrelevant to our discussion.

The gauge field holographically describes the operator $\bar{q} \gamma^\mu q$ 
and its source, a background $U(1)$ gauge field for baryon number. 
We will use
$F_{ab}$ below to introduce a constant magnetic field (eg $F_{12} = -
F_{21} = B/(2\pi\alpha
')$)~\cite{Filev:2007gb} but for the moment keep it zero.

We embed the D7 brane in the $t$, $\vec{x}$, $\rho$ and $ \Omega_3$ directions of
the metric but to allow all possible embeddings must include a
profile $L(\rho)$ at constant $\Omega_1$. The full DBI action we
will consider becomes one dimensional: 
\begin{eqnarray}
  S_{\mathrm{DBI}} = \caln \int \dd t \dd \vec{x} \dd \rho 
  {\cal L}(L,L';\rho) \,, \label{DBI0}
\end{eqnarray}
where  ${\cal N} = N_f T_{D7} 2\pi^2$ and
\begin{equation}
  {\cal L} = - \rho^3 \sqrt{(1+ L'^2)}
    \,.
\end{equation}

The Euler-Lagrange equation for the embedding is then
\begin{equation}
\partial_\rho \left( \rho^3 L' \over  \sqrt{(1+ L'^2)} \right) =0
\end{equation}

At large $\rho$ the classical solution from \eqref{DBI0} behaves as
\begin{equation} \label{asym}
L(\rho) \sim  m + {c \over \rho^2}+ \cdots \,,
\end{equation}
where $m$ is proportional to the quark mass and $c$ to the quark
condensate.

\begin{figure}[]
\centering
\subfigure[]
  {\includegraphics[width=6cm]{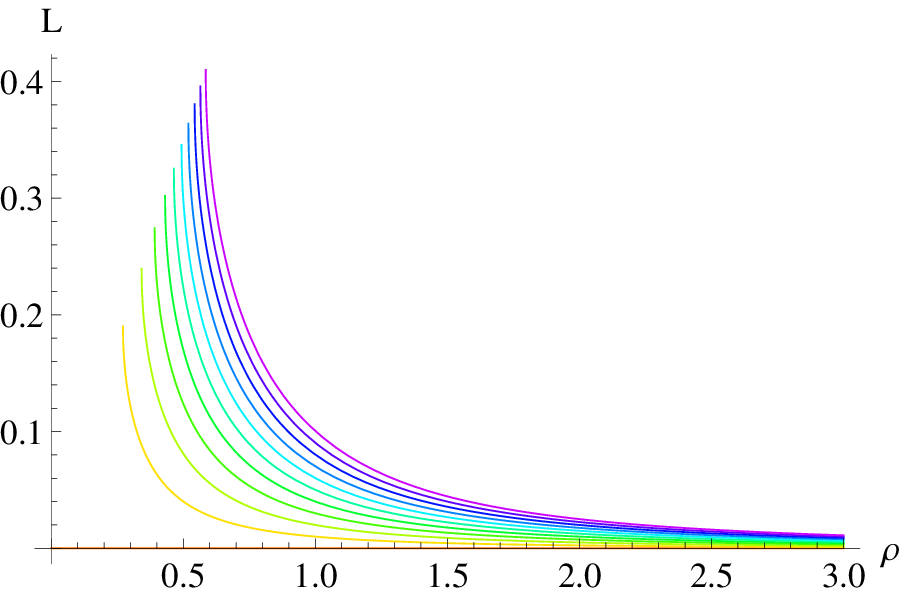}}
\subfigure[]
   {\includegraphics[width=6cm]{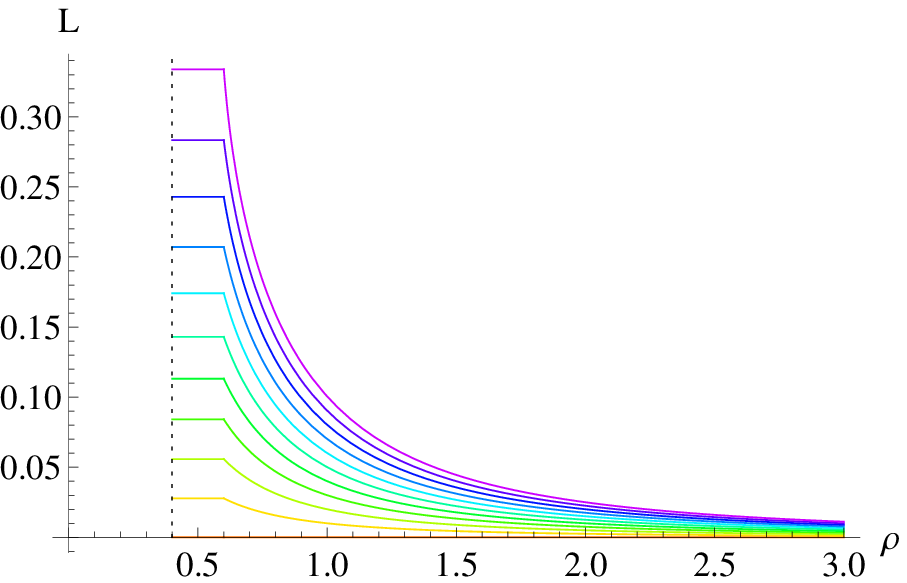}}
  \caption{Plots of D7 embeddings in the ${\cal N}=2$ theory. (a) Full
embedding solutions for $m=0$. (b) The mebddings interrupted by a two scale
cut off. 
           }\label{new1}
\end{figure}

Numerically we can shoot into the IR from a UV solution with particular 
values of $c$ and $m$. For the particular case $m=0$ we show
such flows in Fig. \ref{new1}(a). 
All except the $c=0$ solution appear 
to stop at some finite $\rho$.
In this case we can find the analytic solution to investigate this in more detail.
The real solution valid in $\rho \in (\rho_c,\infty)$, where $\rho_c = (2 c)^{1/3}$ is 
\begin{equation} \label{n2sol}
L(\rho) = m + {c \over \rho^2} {\,}_2\mathrm{F}_1 [1/3,1/2,4/3, 4c^2/\rho^6]
\end{equation}
with
\begin{equation} \label{lp}
L'(\rho) = {  2c \over \sqrt{\rho^6 - 4 c^2 }}
\end{equation}

From (\ref{lp}) we can see that the gradient of the embedding becomes complex at 
$\rho_c = (2 c)^{1/3}$. These results match very well to the numerical
results in Fig \ref{new1}(a) and provide an explicit form for the behaviour at $\rho_c$.
The only solution that survives to $\rho=0$ is the flow with $c=0$.

One would naively like to plot the effective potential $V(c)$
generated by the holographic flows to show the $c=0$ solution is the minimum. 
However, since all but one flow (c=0) become ill-defined this is confusing.
Understanding why there is no IR effective potential provided by holography is one of our goals. We
will adopt the recently suggested idea that we should approach the holographic description
in a Wilsonian manner. In particular we will introduce a cut-off in $\rho$, the holographic 
direction for quark physics (i.e. the radial direction on the D7 brane), which we will call $\epsilon$ 
and study the theory as a function of changing that cut-off.

Thus specifically to convert the ``off-shell'' flows, with non-vacuum values of the quark condensate, into kosher flows
we will interrupt them with a cut-off at an intermediate value of
$\rho$. In the UV we find the Euler Lagrange equation solutions with large $\rho$ asymptotics $c/ \rho^2$
and solve down to the cut-off.  A technical issue arises at this point though. The flows ending 
on the cut-off do not share the same IR boundary conditions since they meet the cut-off at arbitrary
angles. Formally one should not compare their actions in a Euler-Lagrange analysis. 

To cure this let us imagine
a more general structure for our cut-off in which it has finite width. We introduce two cut-offs in
$\rho$, $\epsilon_-$ and $\epsilon_+$. In the UV from infinite $\rho$ down to $\epsilon_+$ we use the flows 
in Fig \ref{new1}(a). Then we match these flows to flows beginning at $\epsilon_-$ with $L'(\epsilon_-)=0$
and ending at $\epsilon_+$ at the same point as the UV flows. We pick this
boundary condition at $\epsilon_-$ because it naturally matches on to the boundary condition of the regular flows
as $\epsilon_- \rightarrow 0$ ie $L'(0)=0$. We show example flows in Fig \ref{new1}(b). Now all of our flows 
have the same IR and UV boundary conditions. 

Having introduced this cut-off structure it is actually
most natural to remove it by taking $\epsilon_+ \rightarrow \epsilon_-$. In this example the
flows between $\epsilon_-$ and $\epsilon_+$ simply become short straight lines whose action
vanishes as the two cut-offs coincide. This digression therefore is just to justify that one can
effectively consider the UV flows down to the common $\epsilon$ to share IR boundary conditions and directly
compare their action sensibly. In other words we assume a small change to the flows at the cut-off
that bends them to satisfy $L'(\epsilon)=0$ but assume this doesn't make a large change to the action.
Here this seems rather trivial but we shall see much more structure emerge in the later example 
with density.  

We now proceed in this case with the single cut-off $\epsilon$. 
For each choice of $\epsilon$ we can plot a potential (density) given by
\begin{equation} 
V_{\mathrm{eff}}(c) = - \int_\epsilon^\infty d\rho \ \label{DefV}
  {\cal L}(\rho)
\end{equation}
which is nothing but a Euclidean on-shell action \eqref{DBI0} normalized by $\caln$ and 
a field theory volume (so it is a density.). These actions diverge in the UV but the difference between
them determines which is preferred (or they can be regulated by 
adding a holographic counter term $\sim \L^4/4$, 
where $\L$ is a UV cut-off to be set to $\infty$ at the end). A minus sign comes from Euclideanization.  
To regulate these flows we will always compute the difference in action from the flat embedding $L=0$
with the equivalent cut-off $\epsilon$. The
$L=0$ embedding will therefore always lie at $V=0$ in our plots. The potential should be viewed as
the potential energy incurred for a particular value of the condensate from scales above the cut-off $\epsilon$.
In other words this potential is the equivalent of the potential in the ``bare'' Lagrangian written down
for the theory at the given cutoff which should be used in conjunction with quantum (or holographic)
behaviour below the cut-off. This is exactly the Wilsonian paradigm. 

\subsection{Wilsonian Potentials}

\begin{figure}[]
\centering
  {\includegraphics[width=8cm]{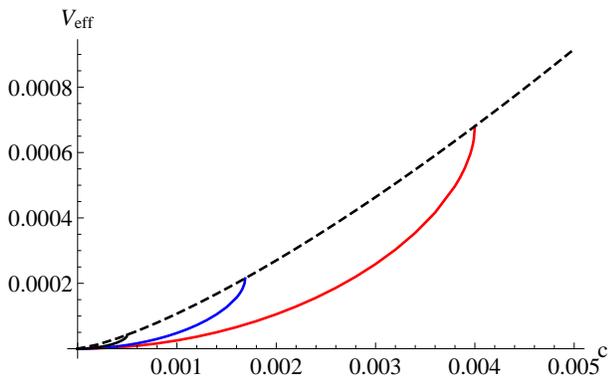}}
  \caption{Plots of the effective potential against the parameter $c$ in the ${\cal N}=2$ theory. Plots for $\epsilon = 0.2, 0.15, 0.1$ (red,blue,black) from right to left. The dotted curve is the extended IR effective potential
with the values of the action frozen at the point the embeddings become complex. }\label{new2}
\end{figure}

Using the methodology described above we can now plot the Wilsonian effective potential
for the quark condensate as a function of cut off scale $\epsilon$. We show this 
in Fig \ref{new2}. Reassuringly $c=0$ is the minimum of the potential at all values
of the cut off as we would expect in the supersymmetric theory. 

The finite extent of the plots in $c$ is a result of solutions
with larger $c$ having gone complex before reaching the cut off  at $\epsilon$. 
We interpret the removal of states with large values of the condensate 
from the effective potential as a sign that these 
states can not be reached with the energy available in the IR theory. This seems
to match well with a Wilsonian approach and explains the degeneracy if $\epsilon \rightarrow 0$.
In the examples below we will simply omit states that are not in the low energy theory in this sense. 
One could though simply freeze the potential value at the point where the embedding becomes complex
and retain that value for lower choices of the cut off. We plot that version of the IR
effective potential for the ${\cal N}=2$ system as the dotted curve in Fig \ref{new2} - again it is minimized at $c=0$.

\section{Wilsonian Flow for a Chiral Condensate}

We will now move on to study more interesting examples of gauge theories
that induce chiral symmetry breaking in the IR. We next look at the ${\cal N}=2$ theory with an
applied magnetic field which induces a chiral condensate ~\cite{Filev:2007gb}.
We introduce the B field through the D7 brane
world volume gauge field in (\ref{dbi1}). We now have
\begin{eqnarray}
  {\cal L} = - \rho^3 \sqrt{(1+ L'^2)}
   \sqrt{\left(1 + \frac{ R^4}{(L^2+\rho^2)^2}B^2 \right)}
    \,.
\end{eqnarray}

At large $\rho$ the asymptotic solution is again given by (\ref{asym}) and we can
again interpret $m,c$ as the quark mass and condensate.
In the absence of $B$ the theory is conformal so it is natural to write 
all dimensionful
parameters in units of $\sqrt{B}$, the intrinsic conformal symmetry breaking scale,
which we do for our numerical work (i.e. put $B=1$).

\begin{figure}[]
\centering
\subfigure[]
  {\includegraphics[width=6cm]{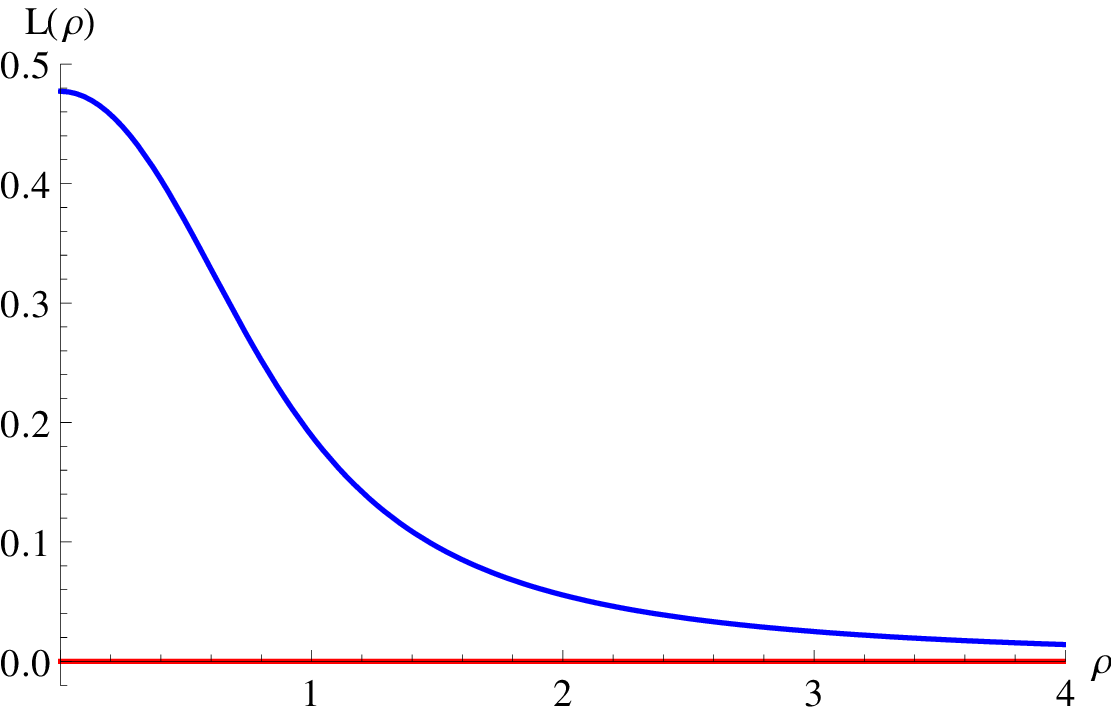}}
\subfigure[]  
  {\includegraphics[width=6cm]{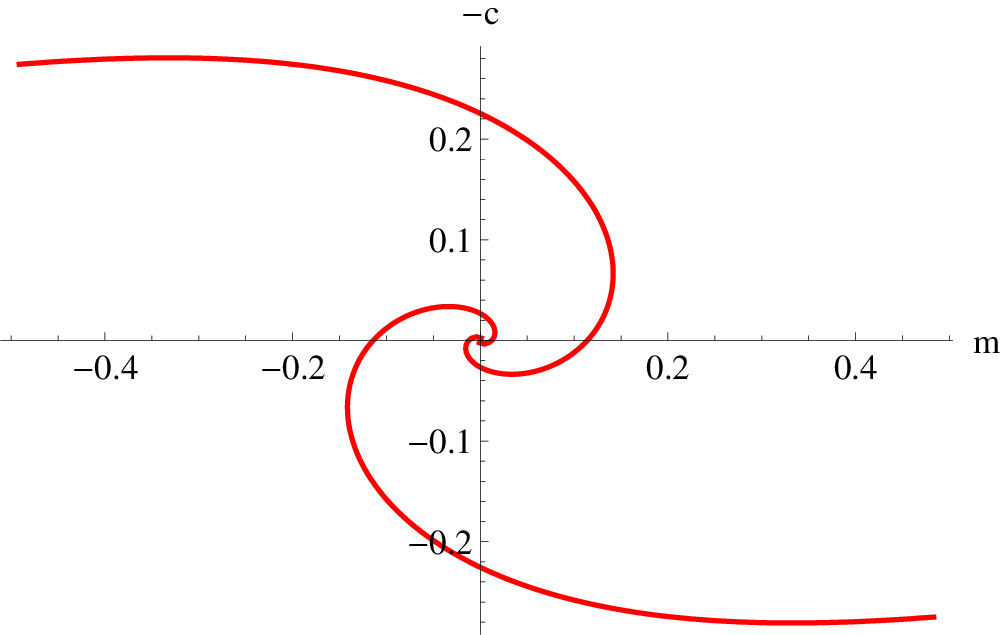}}
  \caption{In (a) we show the flat and vacuum embeddings of the D7 brane probe 
           in AdS$_5\times$S$^5$ with a magnetic field ($B=1$). 
           In (b) we plot the 
           quark condensate against mass extracted from embeddings such as those in (a).
           }
           \label{profiles}
\end{figure}

The solutions of the Euler-Lagrange equations for the embedding $L(\rho)$ are well known~\cite{Filev:2007gb,Evans:2010iy}
and we show two\footnote{In principle there are infinite number
of meta-stable solutions corresponding to the spiral structure in 
Fig \ref{profiles}(b). However, we omit them since 
they are always meta-stable not a ground state.} regular solutions with $m=0$ and with $L'(0)=0$ in Fig \ref{profiles}(a) -
numerically one shoots out from $\rho=0$ to find these. More generally one
can seek such  solutions for any mass $m$ and read off the condensate $c$ from the
large $\rho$ asymptotics. In
Fig \ref{profiles}(b) we show a plot of $c$ vs $m$ for the regular embeddings. It has the spiral
structure discussed in~\cite{Filev:2007gb}. The Fig \ref{profiles}(a) solutions are the $c=0$ flat embedding and the 
largest $c$ solution with $m=0$.
The vacuum energy
of these configurations can be found by integrating minus the holographic action over
the $\rho$ coordinate. These actions diverge in the UV but the difference between
them determines which is preferred (or they can be regulated by 
adding a holographic counter term $\sim \L^4/4 + B^2/2 \log\L$, 
where $\L$ is a UV cut-off to be set $\infty$ at the end).
The curving configuration shown, with the quark condensate, is the
preferred state and the flat embedding is a local maximum of the effective potential.

\subsection{Wilsonian Effective Potentials}

\begin{figure}[]
\centering
\subfigure[]
  {\includegraphics[width=6cm]{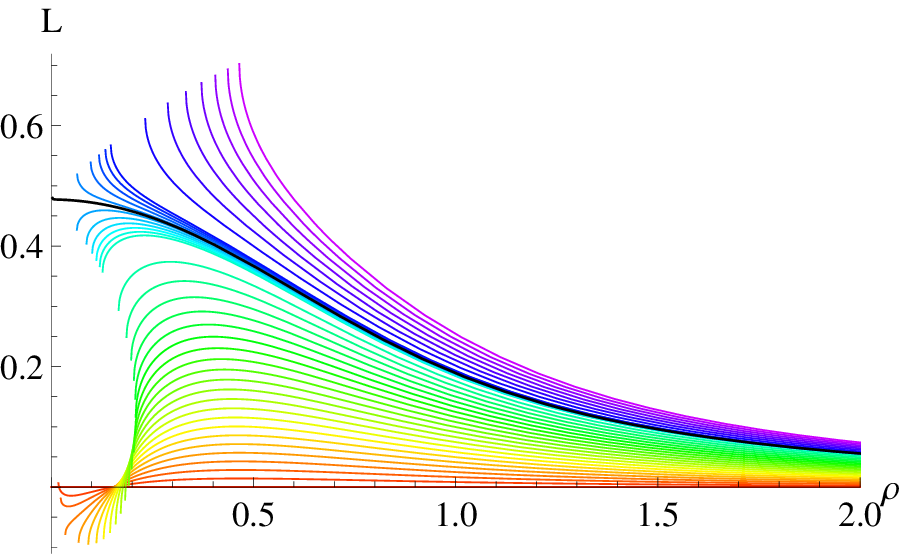}}
\subfigure[]   
   {\includegraphics[width=6cm]{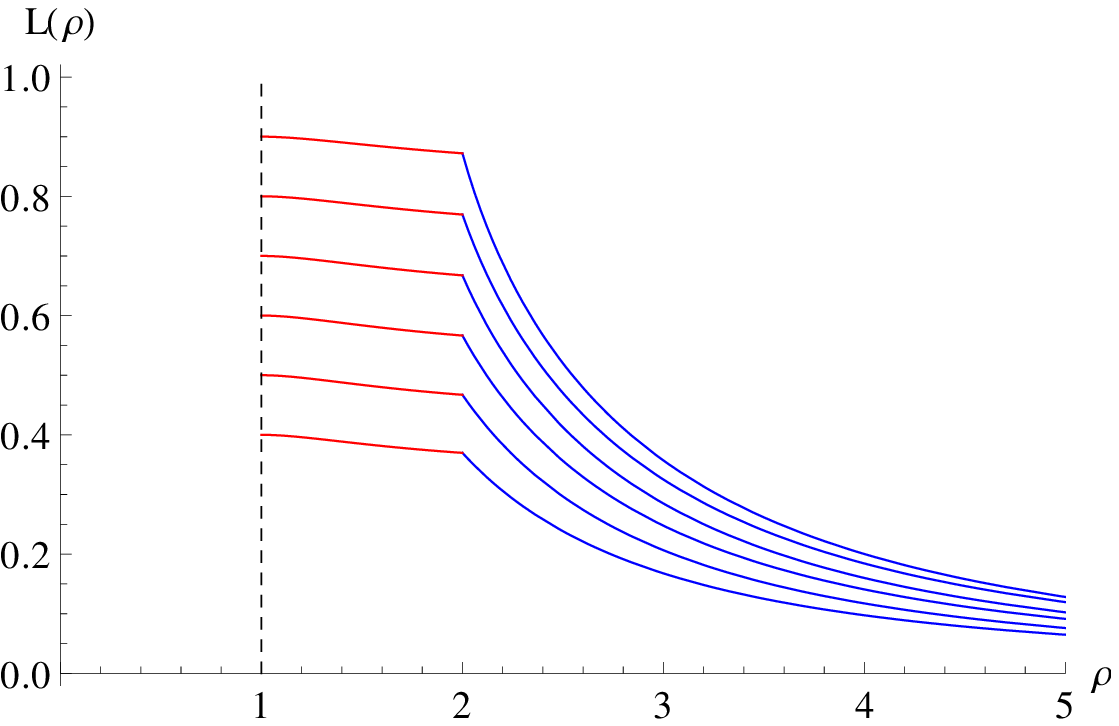}}
  \caption{(a) Plots of the D7 embeddings against $\rho$ for several asymptotic values of the
  quark condensate at zero quark mass with a magnetic field ($B=1$). 
  (b) Those flows interrupted by a 
  two scale cut-off ($\epsilon_-=1$ and $\epsilon_+=2$ here) used to give the flows the same IR boundary condition. 
           }\label{offshell}
\end{figure}

\begin{figure*}[]
\centering
\subfigure[]
  {\includegraphics[width=5cm]{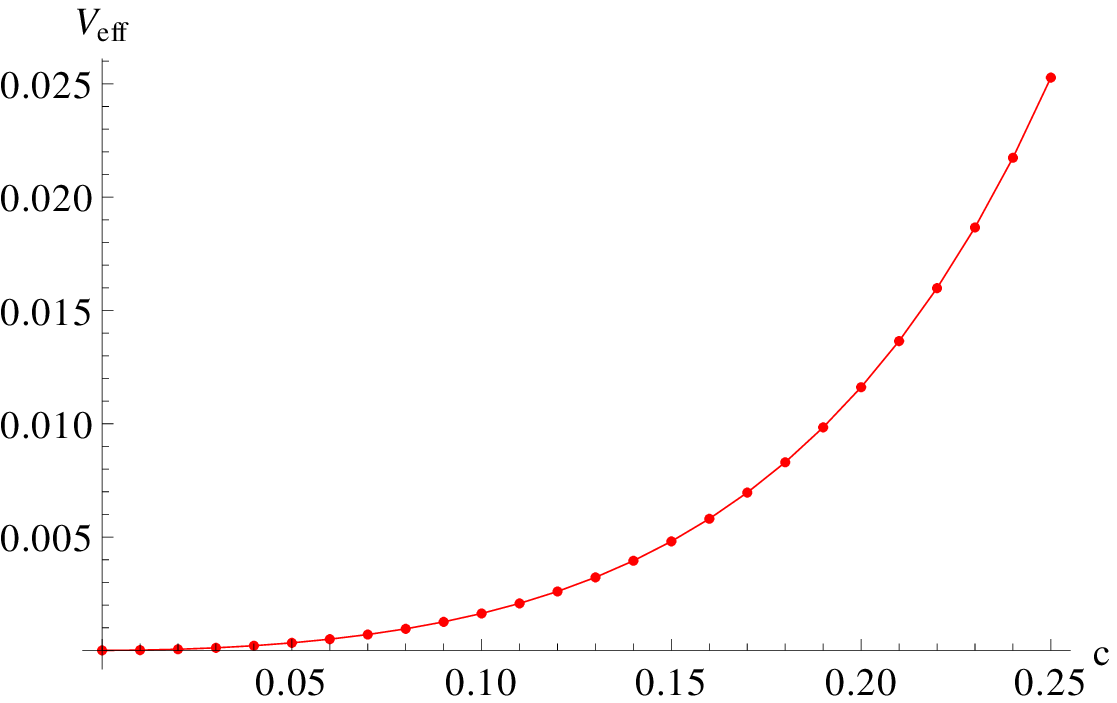}}
\subfigure[]  
   {\includegraphics[width=5cm]{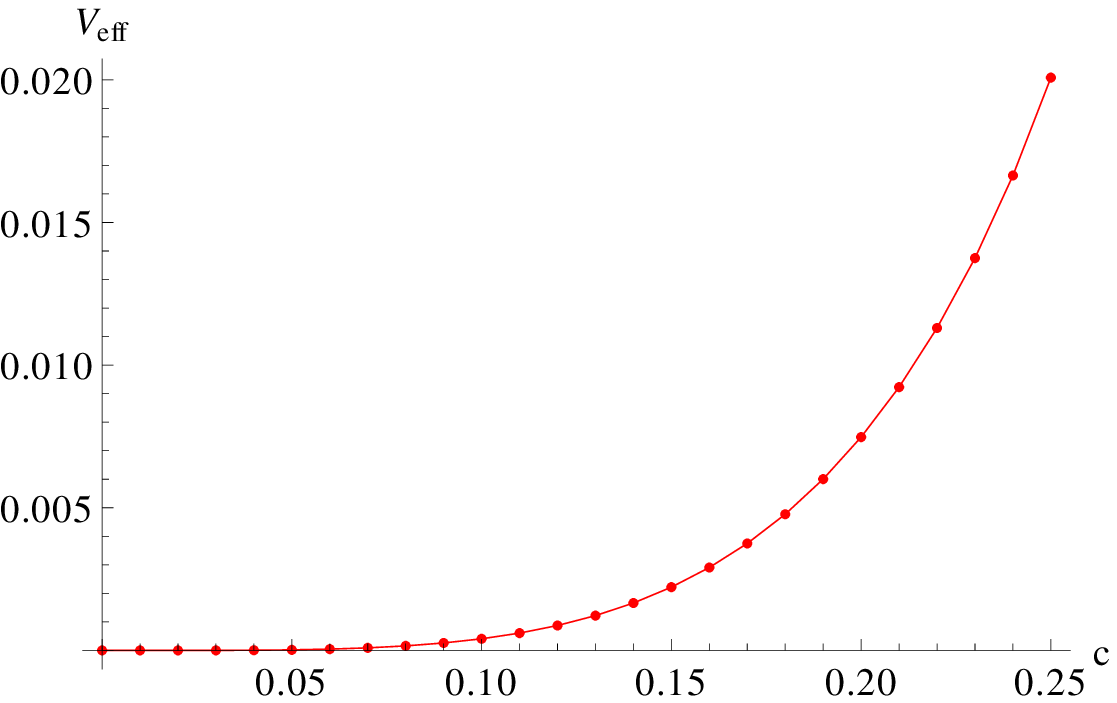}}
\subfigure[]   
   {\includegraphics[width=5cm]{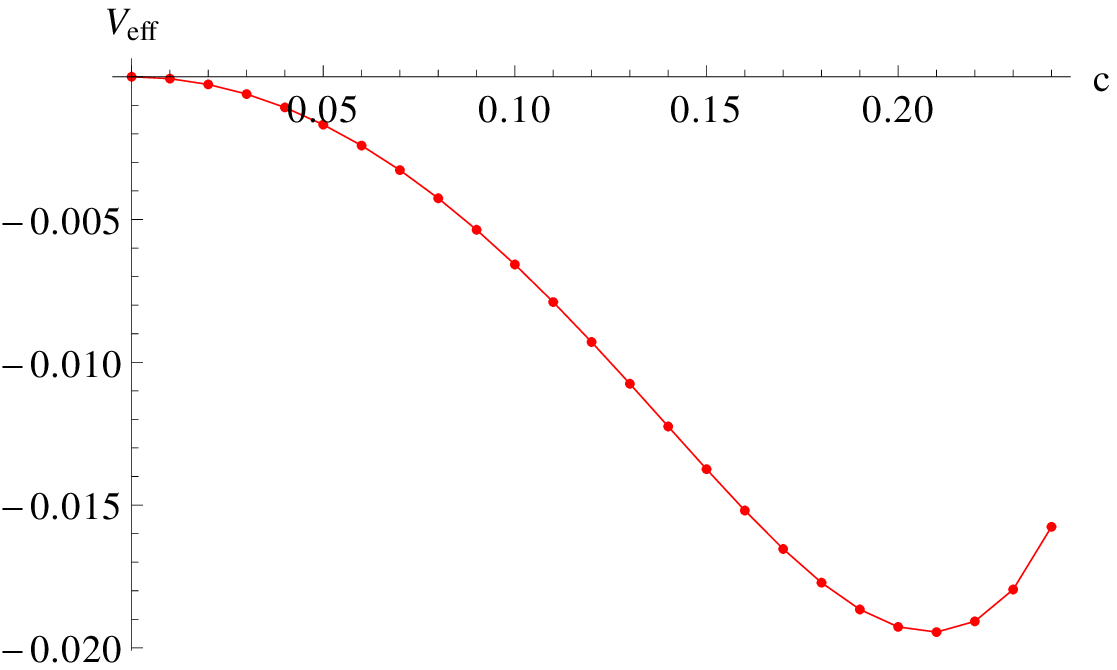}}
\subfigure[]   
   {\includegraphics[width=5cm]{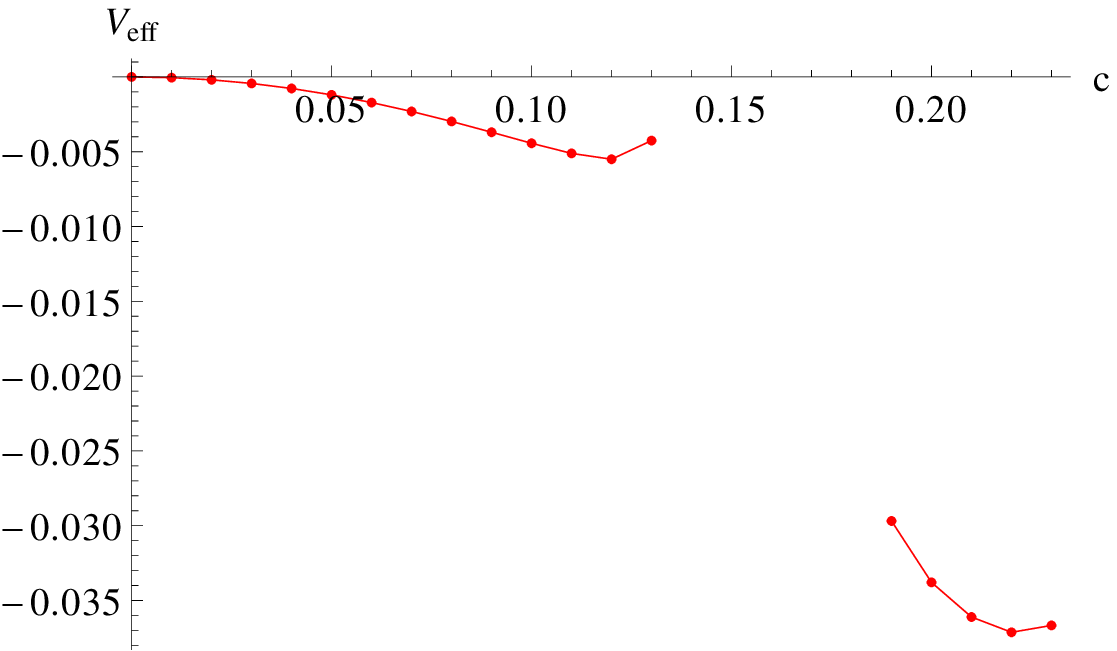}}
\subfigure[]   
   {\includegraphics[width=5cm]{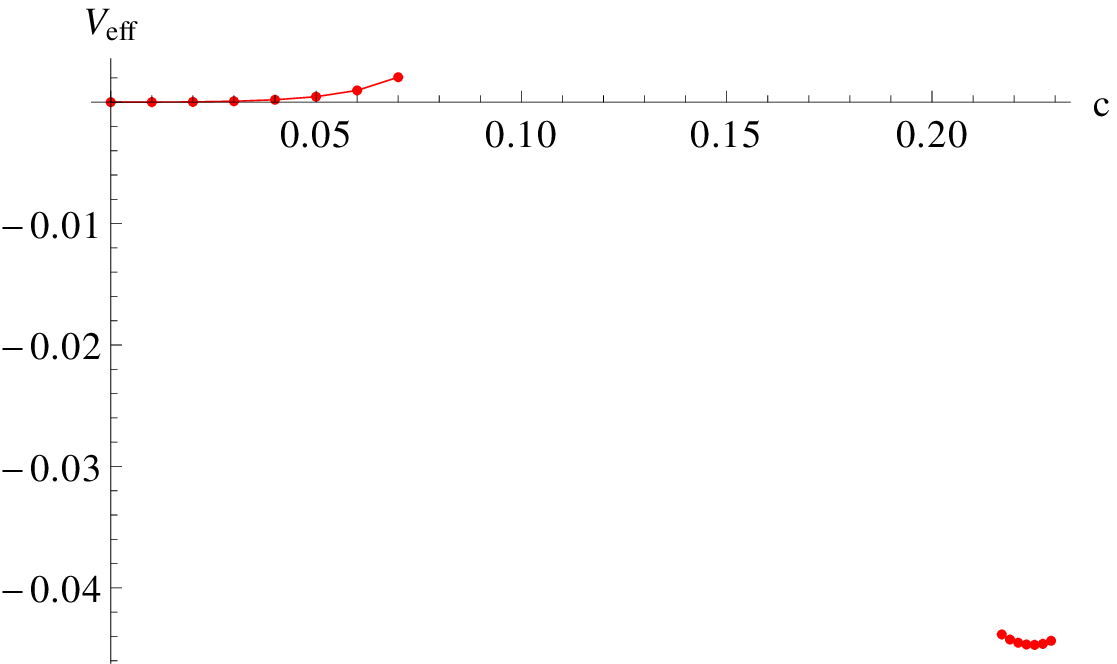}}
\subfigure[]   
   {\includegraphics[width=5cm]{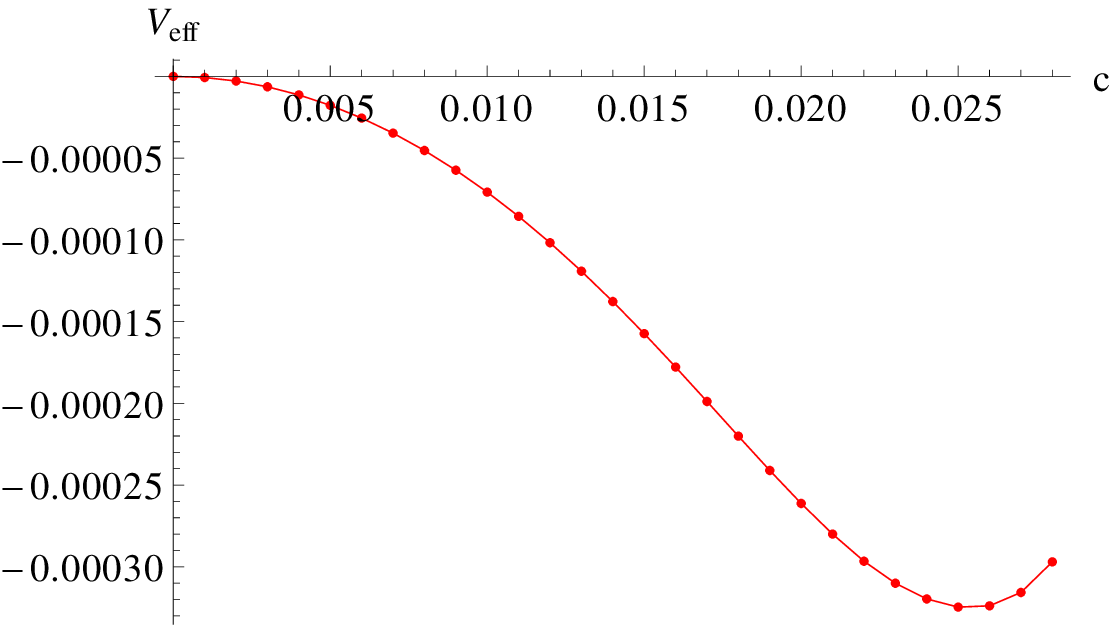}}
  \caption{Plots of the Wilsonian effective potential for the quark condensate in the D3/D7 system 
           with a world volume magnetic field ($B=1$) at different values of the Wilsonian cut-off $\epsilon$
           ($\epsilon = 0.5,0.4686$(phase transition point)$,0.3,0.2,0.15$ and $0.03$  respectively).
           In (a)-(c) we see a second order transition from the unbroken to the broken phase. In
           (d)-(e) we see the IR potential degenerate as non-vacuum states are integrated out. In (f) we show 
           the potential close to the origin at $\epsilon=0.03$ displaying one of the metastable vacua, which corresponds 
           to $c\sim0.025$ also shown in Fig \ref{profiles}(b).
           }\label{Vvse}
\end{figure*}

As in the previous example we would like to plot the effective potential $V(c)$
generated by the holographic flows to show the solutions we have found are the turning points. 
To describe off-shell states we find numerical solutions of the embedding equation
for massless quarks 
that look like $c / \rho^2$ at large $\rho$ and shooting into the interior. 
We plot these flows in Fig \ref{offshell}(a), where it can be seen that most fail to reach the $L$ axis or a deep IR cut-off. We expect that these are associated to the embedding becoming complex by continuity to the $c >> B^{3/2}$ curves, although here we do not have analytic solutions.

To proceed we again introduce a cut off. In 
Fig \ref{offshell}(b) we show such a cut off with the two scale structure, $\epsilon_+, \epsilon_-$ 
that allows us to make the flows all share the same $L'(\epsilon_-)=0$ boundary condition. As
with the pure supersymmetric case we can take $\epsilon_- \rightarrow \epsilon_+$ limit
trivially here, having a single cut off $\epsilon$ at which we end the flows.
We now proceed in this case with the single cut-off $\epsilon$ evaluating the action 
integrated from $\epsilon$ to infinity.

In Fig \ref{Vvse} we plot the potential as a function of $\epsilon$ for the D3/D7 system with magnetic field. 
When $\epsilon$ is large we are describing the UV lagrangian which has no preference for a quark condensate.
As $\epsilon$ decreases we are ``adding in'' more of the strongly coupled quantum effects of the theory 
from lower scales to the bare potential.
The first clear feature shown in Fig \ref{Vvse}(a)-(c)
is that there is a
phase transition at the critical value $\epsilon_c=0.4686 B^{1/2}$ to the chiral symmetry broken phase.
This transition is our first main result. One can think of this transition as being in the spirit
of a Coleman-Weinberg transition - at high energies the theory has no condensate but then strongly coupled loop
effects enter in the IR and break the symmetry. We will return to the deep IR later but let us first 
explore this phase transition in detail.

\begin{figure}[]
\centering
\subfigure[]
  {\includegraphics[width=6cm]{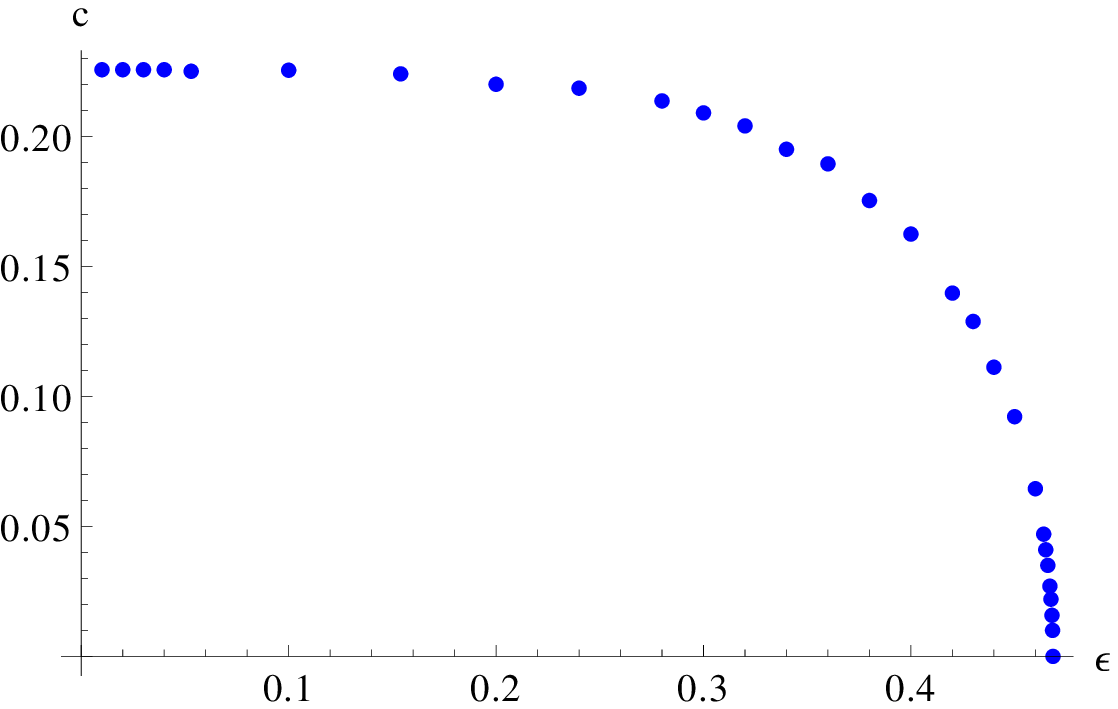}}
\subfigure[]  
  {\includegraphics[width=6cm]{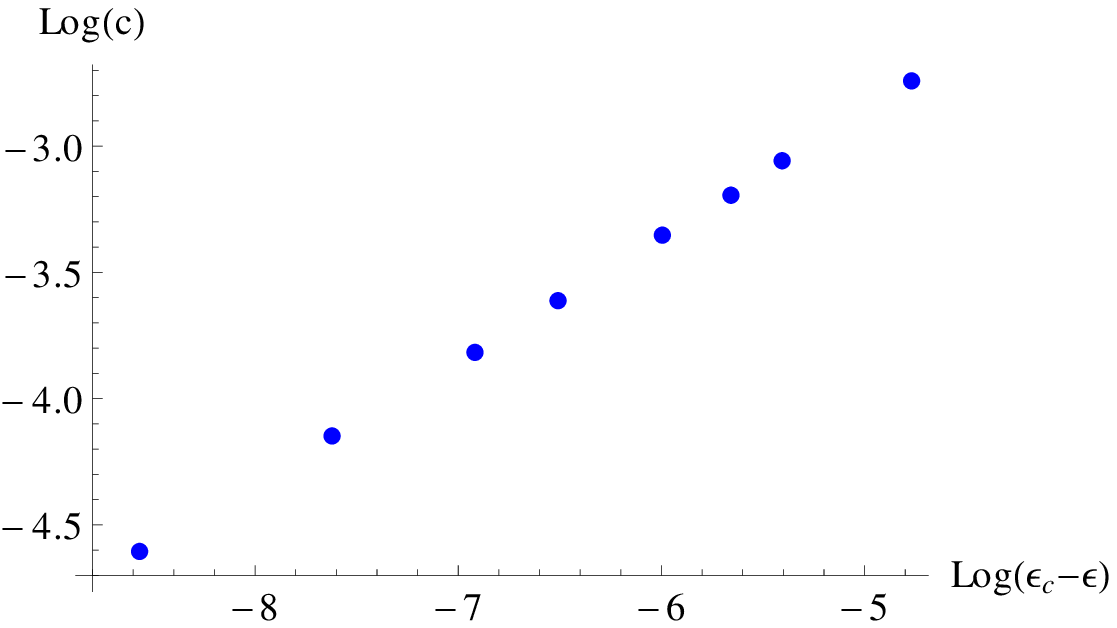}}
  \caption{(a) A plot of the quark condenste $c$ against Wilsonian cut-off scale $\epsilon$. 
   (b) A log log plot of the same close to the transition point to show the mean
     field exponent.
           }\label{cve}
\end{figure}

In Fig \ref{cve} we plot the quark condensate against
$\epsilon$ showing that the transition is second order. We also plot 
$\log c$ vs $\log (\epsilon_c - \epsilon)$ near
the transition point from which we can extract the critical exponent as $1/2$ - the transition is
a mean field one.

In fact close to the transition we can perform a numerical fit to the
potential we have derived of the form
\begin{equation}\label{guess}
  V_\mathrm{eff}(c;\epsilon, B) =  a c^p  + b c^q
\end{equation}
where $a$ and $b$ are functions of $\epsilon$ and $B$. 
Through the range $0.467-0.48$ the fit gives $p=2$ and $q=4$ 
to better than a percent.

\begin{figure}[]
\centering
\subfigure[]
  {\includegraphics[width=6cm]{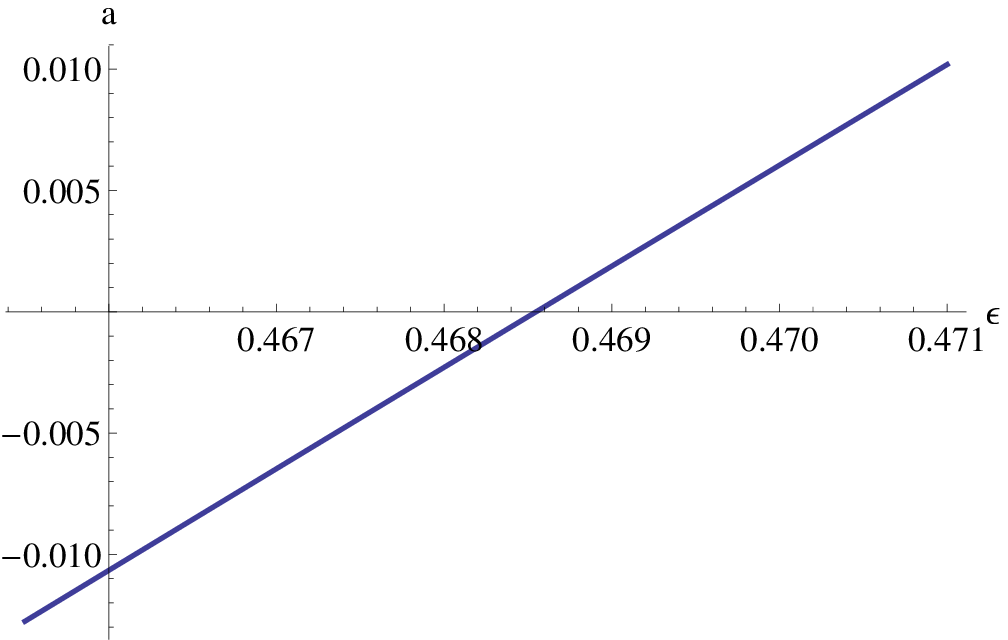}}
\subfigure[]
   {\includegraphics[width=6cm]{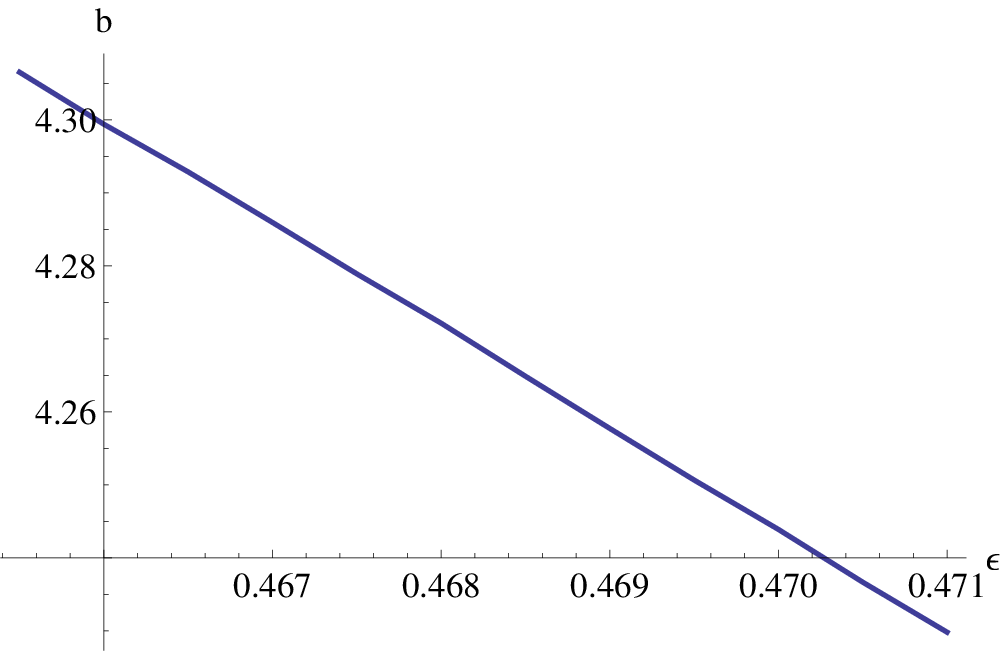}}
  \caption{Plots of the parameters (a) $a$ and (b) $b$ in our fitting potential in (\ref{guess})
           against Wilsonian cut-off $\epsilon$ through the transition point.
           }\label{ab}
\end{figure}

\begin{figure}[]
\centering
\subfigure[]
  {\includegraphics[width=6cm]{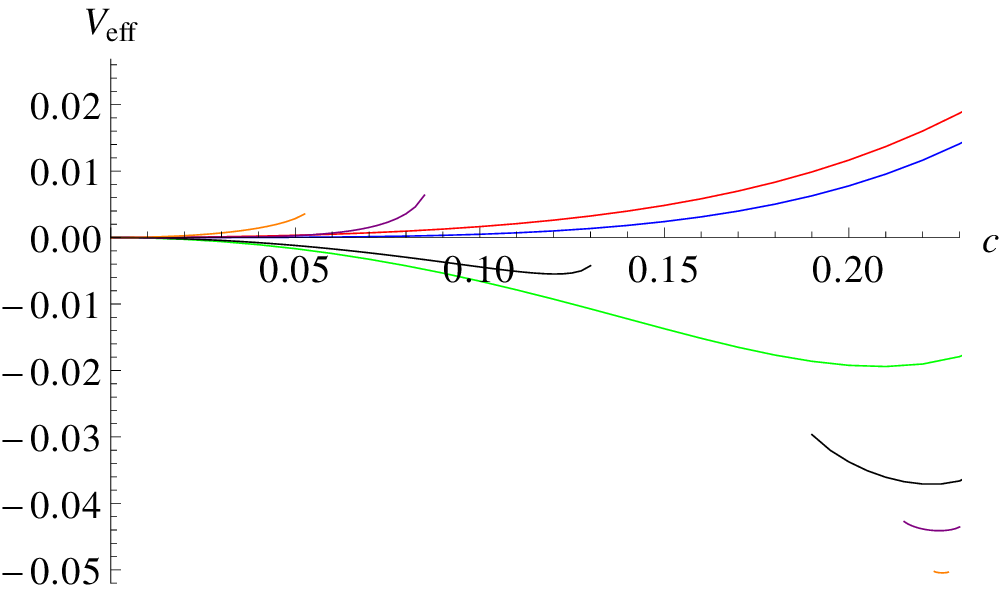}}
\subfigure[]  
   {\includegraphics[width=6cm]{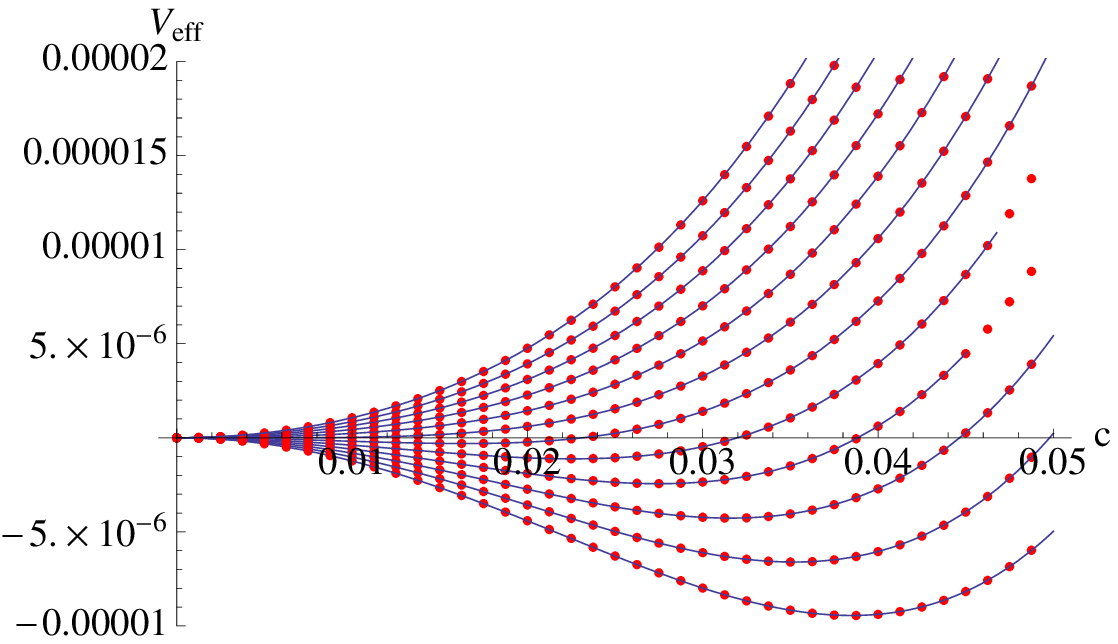}}
  \caption{(a) A summary plot of Fig \ref{Vvse} showing the Wilsonian potential at six different choices of $\epsilon$ on a single 
            plot. (b) The potential and fit (solid line) close to the transition point - the dots are holographically
            determined data.
           }\label{fit}
\end{figure}

The existence of such a potential (which is not present in the far UV theory at all) corresponds to the emergence
of multi-trace operators in the spirit of the discussion in \cite{Heemskerk:2010hk,Faulkner:2010jy}. We next plot
the coefficients $a,b$ against $\epsilon$ in Fig \ref{ab}. 
$a$ and $b$ are approximated as linear functions close to the transition. 
$a$ is proportional to $(\epsilon - \epsilon_c)$ and $b$ is always positive. This is all standard mean field expectations. 
In Fig \ref{fit}(a) we plot the potential for different $\epsilon$ on a single plot. In Fig \ref{fit}(b) we plot the potential for values of $\epsilon$ close to the transition, where the points are
the numerical data whilst the solid curves are our fit potentials.
\begin{equation} \label{V2}
\begin{split}
V_\mathrm{eff}(c;\epsilon, B) & = - 4.17\left(0.4686 - \frac{\epsilon}{\sqrt{B}} \right)
\frac{c^2}{B} \\ 
& \quad + \left(-13.98 \frac{\epsilon}{\sqrt{B}} + 10.81\right)\frac{c^4}{B^4} \ .
\end{split}
\end{equation}

It is clear that the transition is well described by the expected mean field potential. 
We stress though that here we have derived
this form holographically. 

Another key feature is visible in Fig \ref{Vvse}(d)-(e). When we impose a cut-off at
$\rho=\epsilon$ we are excluding some range of condensates. In fact even for large $\epsilon$, flows with very large $c$
become complex in the UV before they reach the cut-off as we saw in the ${\cal N}=2$ supersymmetric theory.
 Qualitatively
it makes sense that if we write a Wilsonian effective model at low energies then large condensate
values should be excluded from the theory since such states would be completely integrated out - they have an
energy above the cut-off. 

As we move to much lower $\epsilon$, the deep IR, in addition some intermediate ranges of condensate 
also disappear from the
effective description in the same fashion. This is why there are breaks in the potential plot
for low $\epsilon$. Again qualitatively
these states have such high energy relative to the low cut-off scale of the effective theory that they
are integrated out.

If $\epsilon$ is strictly taken to zero only the regular flows
corresponding to the turning points of the potential have finite action.
This is clear from
Fig \ref{offshell}(a). The effective potential we are computing therefore degenerates in the deep IR. 
In this Wilsonian language though this degeneration seems entirely
appropriate - when we have integrated out anything above the vacuum energy it is no surprise we are 
left with only the vacuum state in our effective theory.

In addition to the vacuum state we have considered so far, in principle, at $\epsilon = 0$ there are an
infinite number of meta-stable vacua near the flat embedding. This is shown in the self-similarity (spiral) structure in the $c$ vs $m$ plot in Fig \ref{profiles}(b). This self-similarity 
is realized in our Wilsonian context as follows. 
When $\epsilon$ is large $c=0$ is the global minimum. As $\e$ decreases 
it becomes a local maximum and a new ground state forms. This is
shown in Fig \ref{Vvse}(c). As $\e$ decreases more, $c=0$ becomes 
a local minimum Fig \ref{Vvse}(e), which is preparing 
to produce a new meta-stable vacuum. 
As $\e$ decreases yet further then $c=0$ becomes a local maximum again 
Fig \ref{Vvse}(f) leaving 
a meta stable vacuum. This process (from Fig \ref{Vvse}(c) to Fig \ref{Vvse}(f))  will continue as $\e \ra 0$ leaving more and more meta stable points.
The first of these metastable vacua is visible in Fig \ref{Vvse}(f) where we have focused 
near the origin at yet lower $\epsilon$.

Finally here we should comment that the
precise meaning of these phase transitions and absent regions will depend on the precise choice of cut-off. 
The cut-off
in $\rho$ seems natural in the D7 context but without a precise link between the holographic direction and the
field theory RG scale there is some ambiguity. For example one could have chosen to make the cut-off at constant
$r= \sqrt{\rho^2+L^2}$ surface rather than constant $\rho$. Actually we find that particular choice
unnatural because the true IR vacuum state 
would be missing from the $\epsilon=0$ theory since the chiral symmetry breaking flow does not hit $r=0$. 
Further at the point where the true vacuum disappears from the 
IR theory its vacuum energy remains above that of the flat embedding because the flat embedding
has a missing contribution to its energy from where it extends below the cut off - 
we don't see the true vacuum emerge at any scale.
This discussion shows though that the choice of cut-off could be
dependent on the explicit flow. Although this identification remains an outstanding problem we believe
the Wilsonian style description we have presented is very plausible and qualitatively
helpful in understanding the holographic description.

\subsection{Towards an on-shell IR effective potential}

\begin{figure}[]
\centering
  {\includegraphics[width=6cm]{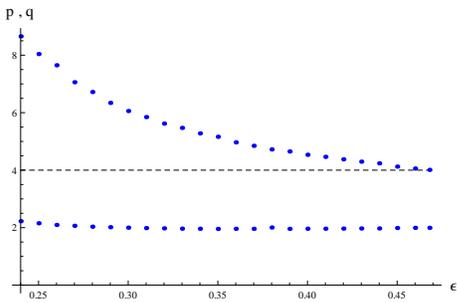}
     }
  \caption{The fitting powers $p,q$ in (\ref{guess}) as a function of Wilsonian cut-off
           $\epsilon$ below the second order transition.} \label{fig:ab}
\end{figure}

It is clear from our Wilsonian analysis above that the deep IR effective potential is highly degenerate.
It is interesting though to track the form of the effective potential below the Wilsonian 
second order transition we described above. We can again fit the potential at varying
$\epsilon$ to a potential of the form in (\ref{guess}). 
In Fig \ref{fig:ab} we show the best fit values of 
the powers $a,b$
with $\epsilon$. Although $a$ stays close to 2, $b$ rises fast at values of $\epsilon$ below the second order
transition. Of course this does not mean that the $c^4$ term is switching off but that 
the coefficients of $c^6, c^8,..$ type terms are
becoming large - one could in principle use a more complex potential fitting form to see this behaviour.
At lower values of $\epsilon$ than those shown in Fig \ref{fig:ab} the potential starts to become degenerate. 

In the deep IR ($\epsilon=0$) only the vacuum configuration with $c = 0.2255 B^{3/2}$ and  $V = -0.05534 B^2$  is 
described holographically. If one wishes one can imagine (using dimensional analysis) 
an effective potential of the form 
\begin{equation}
V_\mathrm{eff} = \alpha \frac{c^2}{B} + \beta \frac{c^4}{B^4} \label{V1}
\end{equation}
and fit for $\alpha$ and $\beta$. We find $\alpha = -2.17659 $ and $\beta=21.4014$.

It is important to stress though that 
the form of this potential is not fixed by the holographic flows - we could have included more terms
with higher powers of $c$ for example that would reproduce the computed values of $c$ and $V$. Also
there is no sense in which this potential is derived away from the minimum -  it is an off-shell
effective potential extrapolated from on-shell values.
That is \eqref{V1} is assumed to be true for off-shell $c$ values so that the
condition  $\frac{\partial V}{\partial c} = 0$  makes sense. 
However, $\a$ and $\b$ are fixed only by on-shell data. 
In our off-shell method at finite $\epsilon$ above, $\a$ and $\b$ are determined by the off-shell data.

The on-shell action also describes a second order phase transition as $B \rightarrow 0$. For $B>0$ the 
quark condensate grows as $B^{3/2}$ as it must on dimensional grounds. It is important to stress the difference
between this transition and the holographic transition we found above with changing $\epsilon$
at fixed B.

\subsection{$B$ and perpendicular $E$}

The pure B field theory is also very closely related to the theory with
both a magnetic and perpendicular electric field present \cite{Karch:2007pd,OBannon:2007in,Erdmenger:2007bn,Albash:2007bq}. The D7 action is

\begin{equation}
  {\cal L} = - \rho^3 \sqrt{(1+ L'^2)}
   \sqrt{\left(1 + \frac{ R^4}{(L^2+\rho^2)^2}(B^2-E^2) \right)}
    \,,
\end{equation}
Clearly this is little different from the previous case since we have an
effective $\tilde{B}= \sqrt{B^2-E^2}$.
Indeed one can think of this system as a boosted version of the static case with
just a $B$ field. However, we find it a useful case to consider in this form because
we can compare the magnitude of the condensate in units of the magnetic field with varying
electric field value. It is interesting to have more than one scale in the problem.

One expects, since $\tilde{B}$ is the only scale in the DBI action that at
small $E$ values
\begin{equation} c \sim (B^2-E^2)^{3/4} \end{equation}
In other words there should be a second order transition at $E=B$ with a non-mean field
exponent.

Above the transition the theory exhibits a singular surface where the DBI action
naively turns complex. This can be resolved by introducing currents induced by the
electric field. The theory becomes a conductor as well as chirally symmetric. We will not
be exploring this high $E$ phase here.

We can study the theory in our Wilsonian approach with a cut-off $\epsilon$. The
theory is equivalent to the analysis of our previous section but with $B$ replaced
by $\tilde{B}$. We can write the effective potential, valid close to
the Wilsonian transition \eqref{V2}
\begin{equation}
{V_\mathrm{eff} \over \tilde{B}^2} = \alpha \left( \frac{\epsilon}{\tilde{B}^{1/2}} - \eta \right) \frac{c^2}{\tilde{B}^3} + b 
\frac{c^4}{\tilde{B}^6}, 
\end{equation}
where
\begin{equation} 
\eta =  \frac{\epsilon_c}{\tilde{B}^{1/2}} = 0.4686,  \quad \alpha = 4.17
\end{equation}

\begin{figure}[]
\centering
\subfigure[]
  {\includegraphics[width=6cm]{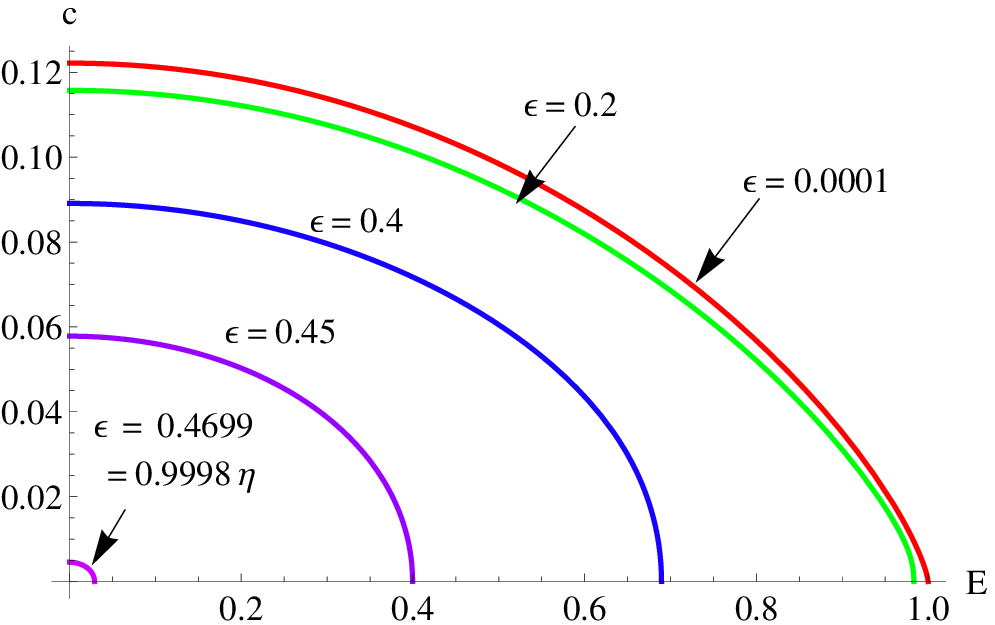}}
\subfigure[] 
  {\includegraphics[width=6cm]{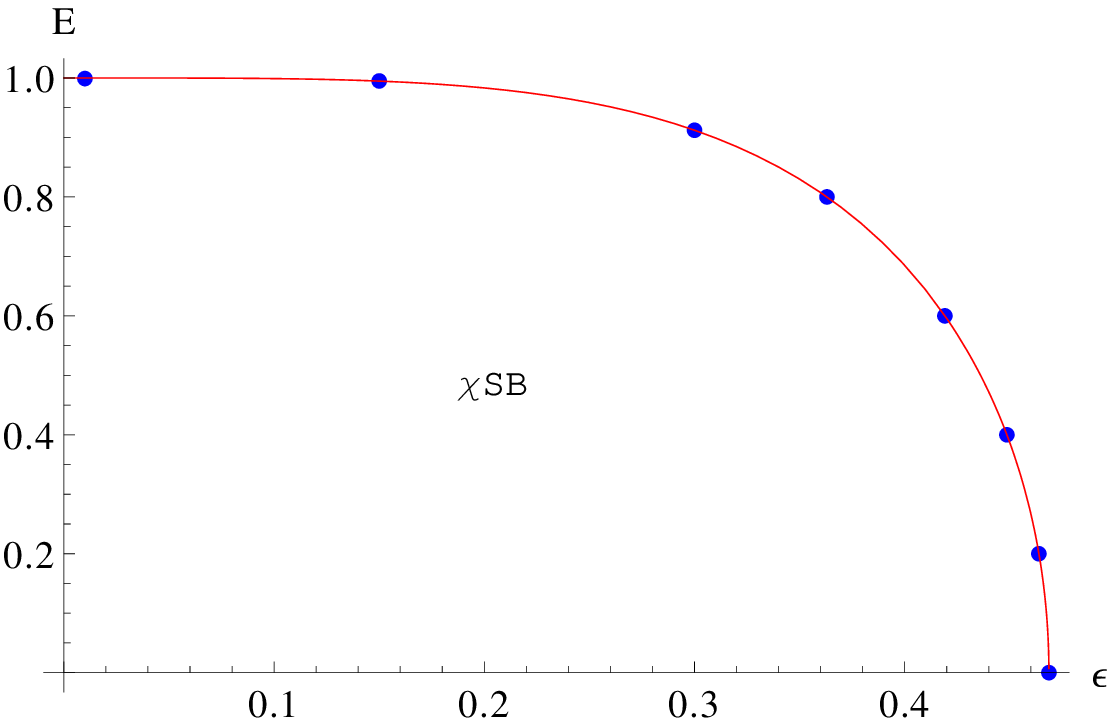}}
  \caption{(a) Plot of the quark condensate against electric field (with a background perpendicular $B$ field ($B=1$))
            at different values of $\epsilon$. (b) Plot of the $E-\epsilon$ phase diagram - the points are derived 
            from the holographic flows whilst the continuous curve is $E= \sqrt{1- (\e/ \eta)^4}$ 
           }\label{fig.cE}
\end{figure}

We can now hold $\epsilon$ fixed and vary $E$.
This potential tells us the full $E$ dependence of the theory near the transition point. In
particular we can determine the transition point from where the mass term changes sign.
\begin{equation} 
E_c^2 = B^2 - \frac{\epsilon^4}{\eta^4} 
\end{equation}
Further by writing $E= E_c + \delta E$ and expanding we find the effective mass squared depends on $E$ as
\begin{equation}
m^2 = \frac{\alpha}{2} \frac{\eta^4}{\epsilon^4} \sqrt{B^2-E^2} E_c (E- E_c)
\end{equation}
In other words the transition is mean field as one moves in $E$ as well as $\epsilon$.

In Fig \ref{fig.cE} we plot the condensate against $E$ for various choices of $\epsilon$ and
the phase diagram in the $E-\epsilon$ plane. It is important to realize that the transition 
for which we have found the effective potential is that at finite positive $B^2-E^2$ and
finite $\epsilon$ close to the transition in $\epsilon$. In other words our effective theory
describes the $c-E$ plot only near the $c=0$ axis. 
We can see that the range of validity of our
effective theory is only away from the point $\tilde{B}=0$ (i.e. away from $E=B$)
\begin{equation} 
\sqrt{B^2 - \frac{\epsilon^4}{\eta^4}} < E < B  
\end{equation}
which clearly has no extent at $\epsilon =0$
The point $\epsilon=0$ on the E axis of the $c-E$ plot
is distinct with a critical exponent of $3/4$ relative to the mean-field exponent
along the rest of the axis. We can not compute the form of the effective potential
for the transition at $\epsilon=0$ other than in the on-shell fashion described in the previous section.

\section{Transitions with $B$-field and Density}

The next models we will explore are the D3/D7 and the  D3/D5 systems with magnetic field, to trigger
chiral symmetry breaking, and density, $d$, which opposes chiral symmetry breaking. 
The phase stucture of these theories has been explored in~\cite{Evans:2010iy,Jensen:2010vd} for the D3/D7 system that
has a second order mean field transtion with increasing density, and in~\cite{Jensen:2010ga,Evans:2010hi} for the D3/D5  
system that displays a holographic BKT transition in which the condensate grows like an
exponential of $-1/\sqrt{d_c-d}$. Our goal is again to use Wilsonian techniques to learn about
these transitions and find the form of the effective potential responsible for the BKT
transitions. This system is more complicated than the pure $B$ system as we shall see 
but we will again enforce that all flows we compare have the same IR boundary conditions
at the Wilsonian cut-off to give a concrete prescription. The outcome 
is a consistent Wilsonian picture of the theories and our ability
to derive the effective potential for the condensate including a potential that
generates the BKT behaviour. We will concentrate first on the D3/D7 system.

\subsection{Density in the D3/D7 system}

\begin{figure}[]
\centering
\subfigure[]
  {\includegraphics[width=6cm]{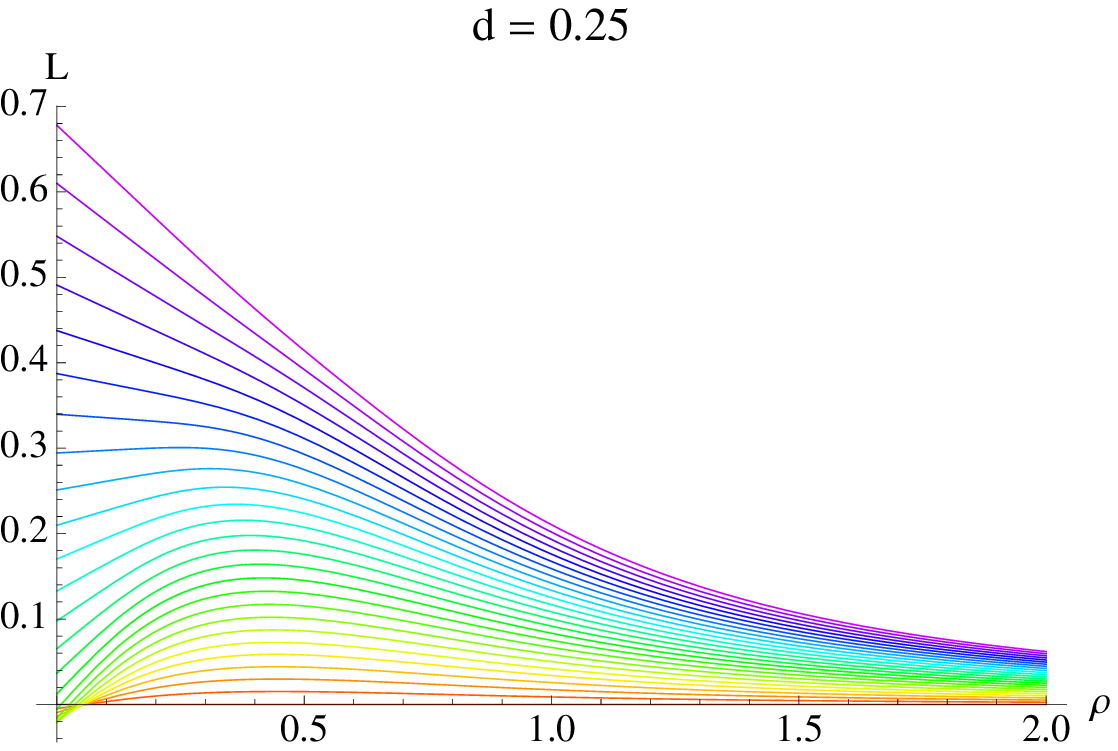}}
\subfigure[]  
  {\includegraphics[width=6cm]{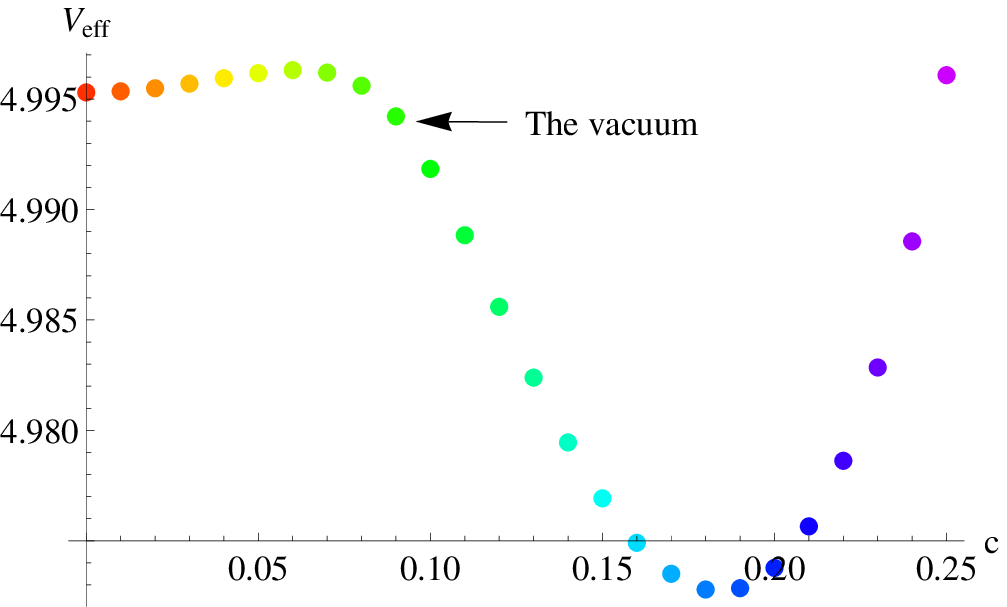}}
  \caption{ (a) D7 embeddings with a fixed $B$ field ($B=1$) and density
  ($d=0.25$) but varying 
             condensate $c$. (b) The incorrect effective potential derived by integrating over the action
             of the flows in (a) - the position of the true vacuum is marked.
           }\label{fig.UV}
\end{figure}

Density is introduced into the theory through a background value for the temporal gauge 
field of the $U(1)$ baryon number \cite{Nakamura:2006xk,Kobayashi:2006sb,Kim:2006gp,Kim:2007zm}. The UV asymptotic form of the field is
$\tilde{A}_t = 2\pi\a' A_t= \mu + d/\rho^2+..$ and describes the chemical potential $\mu$ and the quark density $d$.
The probe D7 DBI action with $\tilde{A}_t$ is given by
\begin{equation}
  \call = -  \rho^3 \sqrt{1+L'^2-\tilde{A}_t'^2}
  \sqrt{1+\frac{B^2}{(\rho^2 + L^2)^2}} \label{Lagd}
\end{equation}
Since the action only depends on the derivative of $\tilde{A}_t$ 
there is a conserved charge density, $d$, defined as
\begin{equation}
  d \equiv \frac{\del\call}{\del \tilde{A}_t'} \,. \label{ddd}
\end{equation}
We may Legendre transform the Lagrangian \eqref{Lagd} to write the action in terms of density
\begin{equation}
\begin{split}
  \call_{\mathrm{LT}} &= 
   \call - \tilde{A}_t' d \\ 
   &= - \sqrt{1+L'^2}
  \sqrt{d^2 + \rho^6 \left(1 +\frac{B^2}{(\rho^2 + L^2)^2}\right)}
  \label{Action1}
\end{split}  
\end{equation}
For fixed $B$ and $d$ we can find solutions to the embedding equation of the D7 brane
with UV behaviour $m + c/ \rho^2$. We plot example flows in Fig \ref{fig.UV}(a). At first sight this 
system seems rather different from the $d=0$ theory - solutions for a large range of
$c$ extended all the way to $\rho=0$. One can naively evaluate the action of these curves and plot it
as an effective potential against $c$ - see Fig \ref{fig.UV}(b). 
This interpretation is though incorrect for
several reasons.

Firstly, these flows all meet the $L$ axis at different angles. This means that they have different IR 
boundary conditions and we should not compare their action in a standard Euler-Lagrange context. 
Further, since $L'(0)\neq0$ these branes are actually kinked at $\rho=0$, when $SO(4)$ rotated to
provide the full D7 embedding.

Secondly, these flows have a non-zero gauge field on their surface that should be sourced. In \cite{Kobayashi:2006sb} the authors
argued that the correct source should be fundamental strings stretched between the D3 (or the origin) 
and the D7 branes. These would be explicitly the quarks corresponding to the density. Such a continuous
distribution of fundamental strings can be absorbed into the D7 world volume and show up as the D7
brane spiking to the origin of the space. The authors of \cite{Kobayashi:2006sb} argued in this way that only the embedding that
ends at the origin was a ``good'' flow and it should represent the vacuum. This is now the standard 
interpretation and has provided a coherent picture across a wide range of problems including density.

Note that, with a naive choice of boundary condition shown in Fig \ref{fig.UV}(a),
the vacuum flow for massless quarks is not 
the minimum of the potential (Fig \ref{fig.UV}(b)).  
This is no surprise since the other flows are not physical.

\begin{figure}[]
\centering
  {\includegraphics[width=6cm]{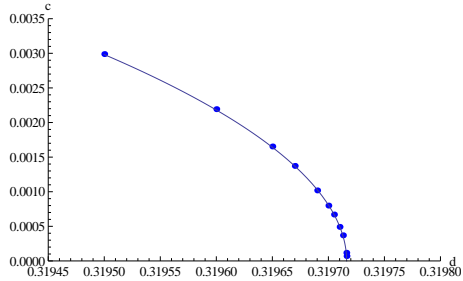}} 
  \caption{A plot of the quark condensate parameter $c$  versus density 
  in the D3/D7 system with a magnetic field $B=1$.}\label{fig.condensate}
\end{figure}

Using the ``good'' flow condition one can compute the quark condensate against density at fixed $B$ (Fig \ref{fig.condensate}). 
There is a phase transition  at $d=d_c=0.3198$. 
It is second order and mean field in nature ($c \sim \sqrt{d_c-d}$).

\subsection{Wilsonian Flows and Potentials}

Our traditional analysis of the D3/D7 system with $B,d$ has again left us with no consistent
supergravity flows that describe off-shell quark condensate configurations. 
Let us try to use a Wilsonian cut-off to provide a hint as to how to proceed. To make progress we will
again introduce a cut-off with structure consisting of two boundaries in $\rho$ at $\epsilon_-$ and
$\epsilon_+$. 

\begin{figure}[]
\centering
  {\includegraphics[width=6cm]{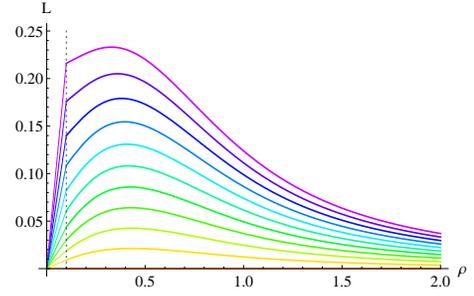}}
  \caption{An example of matching UV flows to flows between the two cut-offs $\epsilon_+$ 
           and $\epsilon_-$ (here $\epsilon_-=0$) which have $L(\epsilon_-=0)=0$.
           }\label{fig.D7e001}
\end{figure}

First consider the case when $\epsilon_-=0$ but $\epsilon_+$ is finite.  For massless quarks, we shoot in from the UV, using a boundary condition of the form $c/ \rho^2$, to the cut-off $\epsilon_+$. These configurations
must be unified to a single IR boundary condition at $\epsilon_-$.  
They also have a non-zero $A_t$ on their world volumes for which we must provide a source. Two of the flows, the
true vacuum and the flat $L=0$ embedding, have smooth extensions to $\epsilon_-=0$ which end at 
the origin. These flows describe good vacuum states of the field theory and must
be included. They, therefore, dictate
what our choice must be for the $\epsilon_-$ boundary condition: we must have the flows satisfying $L(\epsilon_-)=0$, so that we can correctly compare their actions. It is natural then to 
complete the off shell solutions with flows from $L(\epsilon_-)=0$ to $\epsilon_+$ that meet the UV flows.
We show such flows in Fig \ref{fig.D7e001}. 

\begin{figure}[]
\centering  
   {\includegraphics[width=6cm]{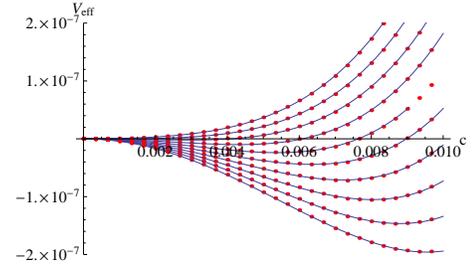}} 
  \caption{The effective potential with $\e_- = 0$ and $\e_+ = 0.01$.  
  $B=1$. Every curve is for a fixed density increasing from bottom to top. The red points are computed from the holographic flows, 
  the blue lines are our fit potential.}\label{fig.D7UVIR}
\end{figure}

We can explore the phase transition with changing density at fixed magnetic field using this cut-off prescription
for the deep IR. In Fig \ref{fig.D7UVIR} we show the effective potential for a range of $d$. Here
we take $\epsilon_-=0$ and $\epsilon_+=0.01$ so the cut-off is very thin. There is
a second order transition as $d$ is raised matching the previously derived critical value of $d$.
Close to the transition point with changing $d$ we can perform a fit to the form of the potential and we find 
\begin{eqnarray} \label{d7potfit}
  && V_\mathrm{eff}(c;d) = -5.53 (0.3198 - d)c^2 + 30.76 c^4 \ , \label{D7fit0}
\end{eqnarray}
This is a mean field potential. In Fig \ref{fig.D7UVIR} the points are holographically derived
data whilst the curves are this fit potential. 
The coefficients of this mean field potential is a 
function of $\e_+$ in general, but a qualitative mean-field potential
form is valid for all $\e_+$ near the phase transition.

Now we would again like to shrink $\epsilon_+$ to $\epsilon_-$ to return to a single cut-off, 
which will fix our potential uniquely as 
a function of $\e = \e_-=\e_+$.  If we do that with
$\epsilon_-=0$ then the UV flows become those we had declared unphysical in Fig \ref{fig.UV}(a) 
but with an added length of D7 extending up the $L$ axis
from the origin. This suggests that that spike is the completion of the flows to make them physical.
One can think of that spike as the fundamental strings that should source the $A_t$ gauge field on
the D7 world volume.

\begin{figure}[]
\centering
  {\includegraphics[width=6cm]{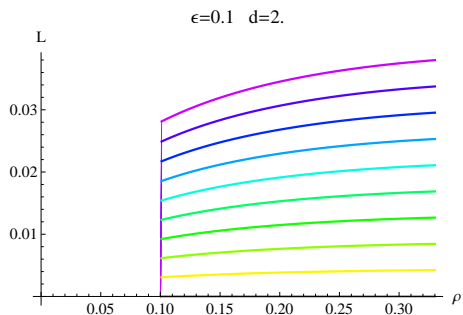}} 
  \caption{Our prescription for completing the flows of Fig \ref{fig.UV}(a) at non-zero Wilsonian cut
           off is to extended them with a spike lying along the cut-off down to the $\rho$ axis 
           as shown.}\label{fig.string}
\end{figure}

What then is the correct way to include the spike contribution at non-zero $\epsilon_-$? 
The natural answer seems to be to maintain the condition $L(\epsilon_-)=0$ 
at finite $\epsilon_-$\footnote{One could just evaluate the sum of the IR and UV contributions of the 
flows in Fig \ref{fig.D7e001} for varying $\epsilon_+$ and 
identifying $\e = \e_+$. If one does so then the vacuum flow
is the minimum of the potential at all $\epsilon$. This picture then does not match our previous  analysis of the pure
$B$ theory - in the UV we expect to find a potential that is minimized at $c=0$ representing the UV bare theory's
unbroken symmetry. We have included too much IR information with such a prescription.}. 
We must enforce the
same boundary condition on all the flows at $\epsilon_-$ and that condition must smoothly map 
to the case $\epsilon_-=0$. When we remove the structured cut-off by
taking $\epsilon_+ \rightarrow \epsilon_- \equiv \e$ we will again be left with a spike from 
the flow of Fig \ref{fig.UV}(a) to the $\rho$ axis, but now
at the scale $\epsilon$ as shown in Fig \ref{fig.string}.

In a Wilsonian sense we would argue this is reasonable since the UV degrees of freedom
should see all the IR physics compressed at the IR cut-off scale $\epsilon$. 
In some sense the UV theory can not distinguish the origin from the point ($L=0, \rho=\epsilon$).
Flows of the form shown in Fig \ref{fig.string}  will then be our cut-off prescription
away from $\epsilon=0$.

The benefits of this configuration are that with a large cut-off the spike simply increases the
action of non-zero $c$ configurations and will leave $c=0$ as the vacuum, whilst in the $\epsilon \rightarrow 0$
limit it will reproduce the known physical solution as the potential minimum. To compute the action of the spike we 
simply take a very thin limit of our two $\epsilon$ prescription - in particular below we will use 
$\epsilon_+ = \epsilon_- + 0.001$.

%
%

Taking this prescription we will now show
that we get a sensible Wilsonian story including a derivation of an appropriate effective potential
for both the D3/D7 and D3/D5 systems.  

\begin{figure*}[]
\centering
\subfigure[$\epsilon = 0.8$. $d = 0.001, 0.1, 0.5, 1$ from bottom.
The theory is always in the chiral phase.]   
   {\includegraphics[width=6cm]{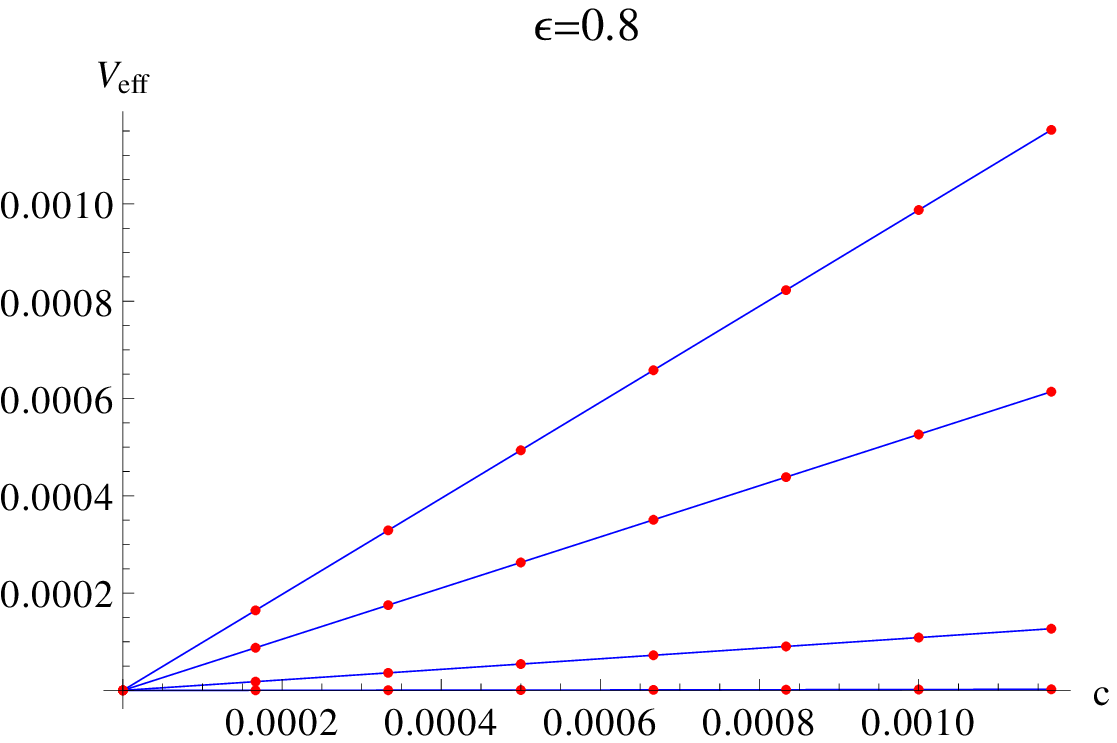}} \ \ \ \ \
\subfigure[$\epsilon = 0.4$. $d=1/300,1/100,1/62$(transition point),$1/30,1/10$ from bottom. There is a first order transition]   
   {\includegraphics[width=6cm]{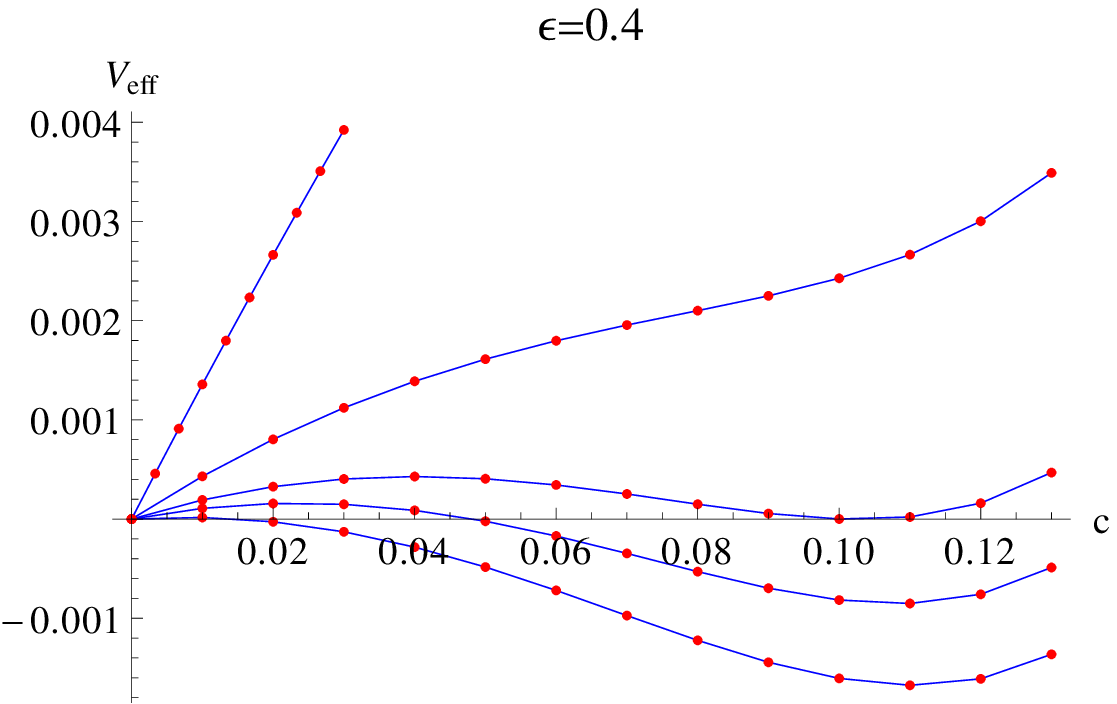}}    
\subfigure[$\epsilon = 0.01$. $d = 0.3085,0.30879$(transition point),$0.309$. There is a second order transition] 
  {\includegraphics[width=6cm]{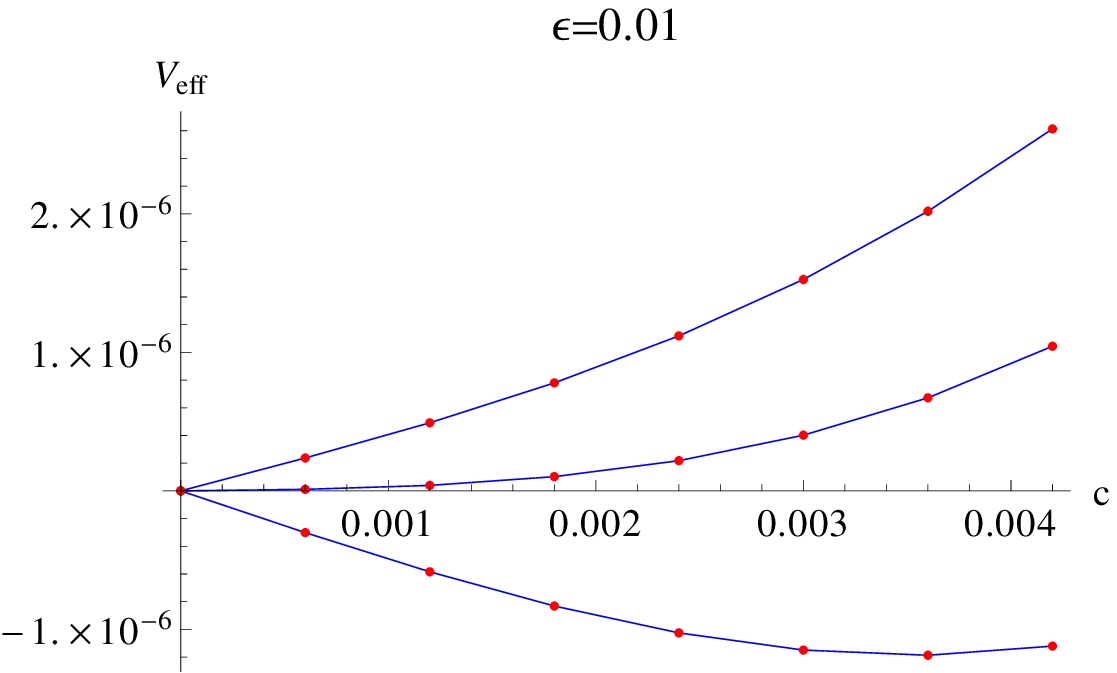}} \ \ \ \ \
\subfigure[The $d$-$\e$ phase diagram.
  The blue colour corresponds to a first order transition and the red to second order.   
   ]
  {\includegraphics[width=6cm]{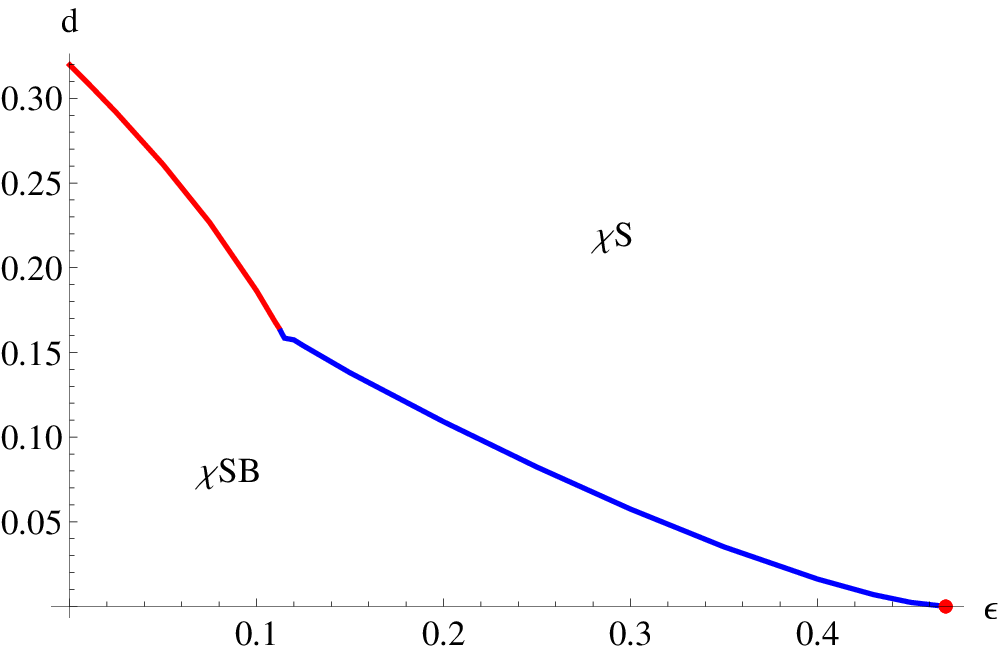}}  
  \caption{ The D3/D7 system $d$-$\e$ phase diagram ($B=1$).
    }\label{fig:D7ed}
\end{figure*}

We first present results for the D3/D7 system.
In Fig \ref{fig:D7ed}(a)-(c) we show the effective potential as a function of $c$ 
for three choices of $\epsilon$ and various $d$. 
$B=1$ is fixed. 
At large $\epsilon$ for all $d$ we see that $c=0$ is the prefered vacuum. As
$\epsilon$  is decreased, provided $d <d_c \sim 0.32$,  there is then a  transition to a chiral symmetry broken phase.
We show examples of values of the cut-off where this transition is first order and second order in 
Fig \ref{fig:D7ed}(b) and (c) respectively.
We can summarize the full picture by drawing the $d-\epsilon$ phase diagram which 
we show in Fig \ref{fig:D7ed}(d). Note that the $d=0$ transition point matches that we found 
above. The $\epsilon=0$ transition point reproduces our potential fit in (\ref{d7potfit}).
The red line is a mean-field second order transition, while
the blue lie is a first order transition.

The insertion of the quark spike with $\epsilon$ has therefore  provided 
a believable Wilsonian picture. In fact though as presented so far the IR diverges from the story
we told at $d=0$. In particular we argued that the potential should degenerate as $\epsilon \rightarrow 0$
as all states other than the vacuum are integrated from the low energy Wilsonian theory. 
We simply don't observe the embeddings that shoot in from the UV 
becoming complex in this system with density for
embeddings with condensate values of order $d$  (very large choices of $c$ do still go 
complex).
We've not been able to find a simple resolution. Most likely the effective 
potential we are deriving here is equivalent to that we produced in the pure ${\cal N}=2$
theory freezing and retaining the potential value at the point where the embeddings go complex - see
Fig \ref{new2}(b). We leave this issue for future thought.

\subsection{BKT transitions in the D3/D5 system}

The D3/D5 system with magnetic field and density displays a (holographic) BKT transition~\cite{Jensen:2010ga}. The
reason it is distinct from the D3/D7 case is that both $B$ and $d$ are dimension 2
in 2+1 dimension and they can be tuned against each other in the deep IR to force the embedding scalar mode of the theory
to violate the Breitenlohner Freedman bound of the effective IR AdS$_2$. The result is that a BKT transition
occurs with an exponential growth of the order parameter (quark condensate)
for $d$ below $d_c = \sqrt{7} B$ ($c \sim e^{-1/\sqrt{d_c-d}}$). This is discussed in detail in \cite{Jensen:2010ga,Evans:2010hi}. Our goal here is to
derive an effective potential for a BKT transition. 

The probe D5 brane is embedded in the $t$ and two $x$ directions of the D3 brane coordinates so that
the quarks live on a 2+1 dimensional defect in the ${\cal N}=4$ gauge theory \cite{Karch:2000gx,DeWolfe:2001pq,Erdmenger:2002ex,Myers:2008me}. The D5 brane also extends in three
directions perpendicular to the D3 brane.
The probe D5 brane DBI action with $\tilde{A}_t$ and $B$ present is given by
\begin{equation}
  \call = -  \rho^2 \sqrt{1+L'^2-\tilde{A}_t'^2}
  \sqrt{1+\frac{B^2}{(\rho^2 + L^2)^2}}
\end{equation}
We may Legendre transform to write the action in terms of density \eqref{ddd}
\begin{equation}
\begin{split}
  \call_\mathrm{LT} &=
  \call - \tilde{A}_t' d \\ 
  &= -  \sqrt{1+L'^2}
  \sqrt{d^2 + \rho^4 \left(1 +\frac{B^2}{(\rho^2 + L^2)^2}\right)} \,.
\end{split}  
\end{equation}
For fixed $B$ and $d$ we can find solutions to the embedding equation of the D7 brane with UV behaviour $m + c/ \rho$.

\begin{figure}[]
\centering  
   {\includegraphics[width=6cm]{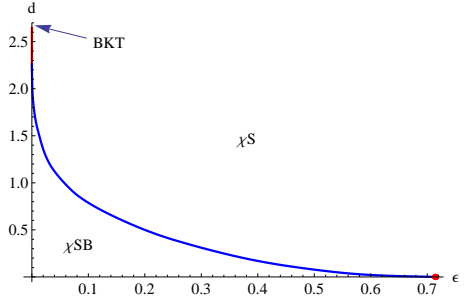}} 
  \caption{The D3/D5 system $d$-$\e$ phase diagram ($B=1$).
 The blue colour corresponds to a first order transition, the red to a second order and the end point BKT transition is labelled.}\label{fig:D5ed}
\end{figure}

Following the last section we will introduce a cut-off at the scale $\epsilon$ and
complete the UV flows with a spike down the cut-off to the $\rho$ axis (we again fix $\epsilon_+=\epsilon_-+0.001$
to generate the spike's action) - see Fig \ref{fig.string}.
We can determine the phase diagram of the theory in the $d-\epsilon$ space at 
fixed $B$ which is shown in Fig \ref{fig:D5ed}. The transition is first order at large $\epsilon$ (the blue line)
then becomes a mean field second order transition at small $\epsilon$ before finally ending on a BKT
transition at $\epsilon=0$. The cut-off behaves like temperature which has already been observed to
convert the BKT transitions to second order with the introduction of any non-zero temperature \cite{Jensen:2010ga,Evans:2010hi}.

\begin{figure}[]
\centering
  {\includegraphics[width=6cm]{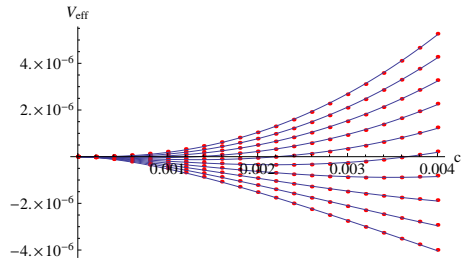}} 
  \caption{The potential with the cut-off $\epsilon=0$
  for the condensate in the D3/D5 system at fixed $B$ ($B=1$) and
  finite density close to the transition. 
   Every curve is for a fixed density increasing from bottom to top.  
   Red points are holographic data and the blue curves are
  the fitted potential in (\ref{d5fit}).}\label{fig.D5UVIR}
\end{figure}

We plot the $\epsilon \rightarrow 0$ potential for varying choices of $d$ in Fig \ref{fig.D5UVIR}.
This potential can not be well fitted by a mean field potential. Instead if we fit to
\begin{equation}
  V_\mathrm{eff} = a c^2 + b c^2 (\log c )^2
\end{equation}
we find a good fit - see Fig \ref{fig.D5UVIR}. The fitting potential in Fig \ref{fig.D5UVIR} is 
\begin{equation} \label{d5fit}
  V_\mathrm{eff}(c;d) = (0.74 d^2 - 4.72 ) c^2 + (0.016 d^2 - 0.12) c^2 (\log c)^2
\end{equation}

This potential form implies that the condensate near the phase transition is
\begin{equation}
  c \sim e^{-\frac{\sqrt{-4a+b}}{2\sqrt{b}}} 
  \sim  e^{\frac{-3.29}{\sqrt{2.77 - d}}}\ . \label{generalc}
\end{equation}
Note that to numerically extract this data one needs to work at extremely high
accuracy near the transition where the condensate is exponentially small. The data shown
are the best we have managed. In
fact the form of the condensate is known analytically to be given by 
~\cite{Jensen:2010ga,Evans:2010hi}
\begin{equation}
  c \sim e^{\frac{-3.86}{\sqrt{2.65-d}}} \label{cth}
\end{equation}
Although we have not reproduced this perfectly our numerics support this form.
We consider it a considerable success to
have derived the potential for the BKT transition in the holographic setting.

\section{Conclusions}

We have analyzed the phase structure of a number of theories that break chiral symmetry 
and have a holographic dual using a Wilsonian cut-off. Including a cut-off allows us
to consider off-shell states, i.e. configurations with a value of the quark condensate
different from that in the true vacuum. We believe that the results give an
improved intuitive understanding of the holographic description and we have been able
to derive low energy effective actions for the phase transitions in these models including
a potential for a BKT transition. The precise identification of our cut-offs in the holographic 
description and the equivalent cut-off structure in the gauge theory remains inexact 
but the spirit seems correct.

We first studied the D3/D7 system (the ${\cal N}=4$ gauge theory with quark 
hypermultiplets in 3+1d). This theory has ${\cal N}=2$ supersymmetry and does not
generate a quark condensate. We nevertheless can in principle plot an effective potential 
for the condensate that should be minimized at zero. The D7 embedding encodes the 
quark condensate and in Fig \ref{new1}(a) we display the Euler Lagrange solutions for the theory 
with different condensate values. In Fig \ref{new1}(b) we insert a cut-off at a finite value of $\rho$ -
here we give the cut-off some finite width and use that width to match all of the solutions
to solutions of the Euler Lagrange equations with the same IR boundary condition at
the lower cut-off. In this case as one shrinks the width of the cut-off, the UV action returns
to that of just the UV flow. In Fig \ref{new2} we plot the Wilsonian potential, evaluated from the
action of the D7 brane above the cut-off, in the theory
as a function of cut-off value. The Wilsonian potential is indeed minimized at zero
condensate at all energy scales. There is a finite extent of the potential in the quark
condensate because the embeddings become complex (which we showed analytically in (\ref{n2sol}).
We interpret the removal of large condensate configurations from the low energy effective actions
as representing those configurations having too high energy to appear in the low energy theory.

The same system with an applied magnetic field has chiral
symmetry breaking. We display the Euler Lagrange solutions for the D7 branes
with different condensate values in Fig \ref{offshell}(a). We introduced a cut off at finite
$\rho$ as in the pure ${\cal N}=2$ theory. 
In Fig \ref{Vvse} we plot the resulting Wilsonian potential as a function of cut-off value. 
The UV of the theory preserves chiral symmetry.
The system then shows a second order mean field transition
to the broken phase at intermediate cut-offs in the spirit of a Coleman Weinberg transition. 
The form of the potential near the transition can be extracted numerically and is displayed in (\ref{V2}).
Finally in the deep IR non-vacuum configurations begin to be integrated out of the 
IR effective theory because they can not be accessed with the IR theory's energy and the 
effective potential again degenerates. 

We translated these results to the theory with a magnetic field and 
perpendicular electric field. This theory shares the same DBI action as the pure B theory
but our results enlarge to describe the phase diagram in the electric field versus cut-off
plane, which we show in Fig \ref{fig.cE}.

We then added a constant quark density into the D3/D7 theory with magnetic field. The
density opposes chiral symmetry breaking. In Fig \ref{fig.UV} we show the D7 embeddings for different
values of the quark condensate which are again ill determined because they end at the IR
axis in a kinked configuration. The true vacuum is the embedding that ends at the origin.
Introducing a cut-off with width as in Fig \ref{fig.D7e001} allows us to define non-vacuum configurations that 
all end at the origin. If we take that cut-off to zero then the embeddings are completed by a spike 
to the origin. For a generic value of the Wilsonian cut-off we have suggested that such a spike
should be introduced along the cut-off as shown in Fig \ref{fig.string}.  Using this prescription we have determined
the phase structure in the density, cut-off plane at fixed magnetic field, see Fig \ref{fig:D7ed}. Here we see first order transitions
with changing Wilsonian scale as well as mean field cases. The IR transitions are mean field 
and the effective potential is of the form given in (\ref{d7potfit}) (in units of $B$).

Finally we looked at the D3/D5 system describing the ${\cal N}=4$, 3+1 dimensional theory with quarks introduced
on a 2+1 dimensional defect. Here with a magnetic field and density the zero temperature theory exhibits
a holographic BKT transition. We have again determined the phase diagram in the density
cut-off plane at fixed magnetic field in Fig \ref{fig:D5ed}. As the cut-off is lowered the transition 
changes from first to second order before becoming a BKT transition in the deep IR. We
have been able to derive an effective potential for the BKT transition in the IR given by
(\ref{d5fit}).

The Wilsonian style analysis therefore allows one to see strongly coupled versions of Coleman Weinberg
like symmetry breaking transitions. It also allows us to derive the low energy effective action
in these theories by defining off-shell configurations. The effective potential for the BKT transition 
is a new result derived here. 

A number of problems remain to be analyzed in these settings including introducing finite temperature
and looking at the non-mean field transitions that lie between mean field and BKT transitions 
\cite{Evans:2010np}. We hope
to study these in the future. These methods will hopefully also be of use away from the probe limit 
where even simple deformations of AdS are typically singular and hard to interpret
\cite{Girardello:1999bd,Gubser:2000nd}. 

\acknowledgments

NE is grateful to the support of an STFC rolling grant and the ESF Holograv collaboration.
KK and MM are grateful for support by the University of Southampton. We thank
Andy O'Bannon and Jonathan Shock for discussions.

\providecommand{\href}[2]{#2}\begingroup\raggedright

\endgroup

\end{document}